%% file: decentralized_beam_codebook_learning_arXiv.tex
\documentclass[onecolumn,draftclsnofoot,12pt]{IEEEtran}
\usepackage{amssymb}
\usepackage{amsfonts}

\usepackage{enumerate}

\usepackage{enumerate}
\usepackage{amsmath,amsthm}
\usepackage{mathtools}
\usepackage{algorithm,algorithmic}
\usepackage{float}
\usepackage{hyperref}
\usepackage{color}
\usepackage{makeidx}
\usepackage{bbm}
\usepackage{graphicx}
\usepackage{lipsum}
\usepackage{soul}
\usepackage{tabularx}
\usepackage{dsfont}
\usepackage[table,xcdraw]{xcolor}

\input{input.tex}
\usepackage{epstopdf}

\newcommand{\sref}[1]{{Section}~\ref{#1}}
\newcommand{\fref}[1]{{Fig.}~\ref{#1}}

\newcommand{\tref}[1]{{Table}~\ref{#1}}
\newcommand{\apref}[1]{{Appendix}~\ref{#1}}
\newcommand{\pref}[1]{{Proposition}~\ref{#1}}

\DeclareMathOperator*{\argmax}{arg\,max}

\newtheorem{prop}{Proposition}

\newcommand{\highlight}[1]{\textcolor[rgb]{0.00,0.00,0.00}{#1}}
\newcommand{\update}[1]{\textcolor[rgb]{0.00,0.00,0.00}{#1}}

\def\x{10}

\newcommand{\subto}{\operatorname{s.t.}}

\begin{document}
\title{Decentralized Interference-Aware Codebook Learning in Millimeter Wave MIMO Systems}
\author{Yu Zhang and Ahmed Alkhateeb \thanks{Yu Zhang and Ahmed Alkhateeb are with Arizona State University (Email: y.zhang, alkhateeb@asu.edu). This work is supported by the National Science Foundation under Grant No. 1923676.}}
\maketitle

\begin{abstract}
Beam codebooks are integral components of the future millimeter wave (mmWave) multiple input multiple output (MIMO) system to relax the reliance on the instantaneous channel state information (CSI). The design of these codebooks, therefore, becomes one of the fundamental problems for these systems, and the well-designed codebooks play key roles in enabling efficient and reliable communications. Prior work has primarily focused on the codebook learning problem within a single cell/network and under stationary interference. In this work, we generalize the interference-aware codebook learning problem to networks with multiple cells/basestations. One of the key differences compared to the single-cell codebook learning problem is that the underlying environment becomes non-stationary, as the behavior of one base station will influence the learning of the others. Moreover, to encompass some of the challenging scenarios, information exchange between the different learning nodes is not allowed, which leads to a fully decentralized system with significantly increased learning difficulties. To tackle the non-stationarity, the averaging of the measurements is used to estimate the interference nulling performance of a particular beam, based on which a decision rule is provided. Furthermore, we theoretically justify the adoption of such estimator and prove that it is a sufficient statistic for the underlying quantity of interest in an asymptotic sense. Finally, a novel reward function based on averaging is proposed to fully decouple the learning of the multiple agents running at different nodes. Simulation results show that the developed solution is capable of learning well-shaped codebook patterns for different networks that significantly suppress the interference without information exchange, highlighting a promising practical codebook learning solution for dense and fast deployment in future mmWave networks.

\end{abstract}

\section{Introduction} \label{intro}
Beam codebooks are essential component in millimeter wave (mmWave) multiple input multiple output (MIMO) system for establishing efficient and reliable links, thereby supporting the stringent communication requirements \cite{Hur2013,Ayach2014,Alkhateeb2014,Alkhateeb2014MIMO,Giordani2019,Heng2021}.
Conventional codebooks, such as the discrete Fourier transform (DFT) based codebooks, include beam patterns that cover the whole angular space in order to provide a full coverage in various deployments \cite{Hur2013,Ayach2014}.
To overcome some of the key drawbacks of the conventional codebooks, site-specific codebook learning is promising in tailoring the beams to suit a specific deployment, hence achieving environment awareness.
One of the major shortcomings of the existing codebook learning solutions, however, is that they are developed by assuming a stationary interference environment, that is, the behavior of the interferer is fixed over time.
Such assumption becomes fragile as the future network becomes denser and denser, since the operations of the adjacent networks might create significant non-stationarity.
This poses challenges to the existing site-specific codebook learning approaches such as unstable convergence and increased learning delay, etc \cite{Shayegan2017}.

Developing codebook learning approaches that are able to deal with the potential interference dynamics is a crucial yet challenging problem, and can be regarded as a more generalized codebook learning task.
One feasible way to tackle the non-stationarity is by building a centralized processing unit that has access and control over the full state of the system. This, however, is normally associated with a few noticeable drawbacks that prevent its adoption in practice. First, it requires large information exchange overhead from and to different base stations, which, for instance, increases the load of the fronthaul and backhaul links. Second, it puts strict synchronization requirements to the networks to ensure coherent measurements.
Third, if the networks belong to different operators, it will also mandate certain information sharing between them.
To address these drawbacks, it calls for solutions that can operate independently at each network, with, ideally, no information exchange between different networks. Such fully decentralized deployment is also promising in reducing the overall algorithm complexity and in improving the scalability to larger networks.
Therefore, in this paper, we focus on developing a fully decentralized codebook learning solution that can learn codebooks at different network nodes under non-stationary interference.

\textbf{Prior work:} 
Beam codebook design for MIMO systems has been an important research topic and has drawn great attentions from both academia and industry \cite{Kutty2016}. The problem has been investigated from various perspectives and for systems with different hardware architectures and operating constraints.
In \cite{Love2003Grassmannian}, the authors study the codebook design under the independent and identically distributed (i.i.d.) Rayleigh fading channels, while in \cite{Xiao2016,Song2017}, the channel model being considered is geometric and is a combination of both line-of-sight (LOS) and non-line-of-sight (NLOS) paths.
Despite its theoretical tractability, the channel distributions in practice can hardly be characterized analytically, and are varying from site to site. As a result, site-specific training can produce higher quality codebook for the given deployment \cite{Alkhateeb2016Frequency,Zhang2020Learning,Bhogi2020,Alrabeiah2022}.
These approaches, however, generally require collecting large offline channel datasets, which consume extensive system resources and interrupt the normal operations. To circumvent the need of full dimensional (i.e., per-antenna) channel datasets, reinforcement learning (RL) based codebook design solutions have been proposed to learn the full codebook without explicitly estimating the channels \cite{Zhang2022Reinforcement,Zhang2023Online}.

In addition to the network side codebook design, existing work also includes designing codebooks for the user equipments \cite{Mo2019,Alammouri2019}. Specifically, how to design effective beam codebook at the mobile device side is crucial for terminals with beamforming capabilities in boosting the performance. In \cite{Mo2019,Alammouri2019}, the authors investigate such problem for the mobile handsets, where a data-driven method is proposed in \cite{Mo2019} to account for practical and irregular antenna radiation patterns, while in \cite{Alammouri2019}, the impact of hand grip on the performance and codebook design is studied.
To reduce the codebook design cost, the possibility of using other information (i.e., non wireless) is also being studied.
This aligns with the general trend of sensing-communication synergy as communication moving towards high frequencies.
In \cite{Chen2023}, for instance, the computer vision is leveraged in identifying the LOS and NLOS user grids to efficiently construct the codebook beams, hence achieving user distribution awareness.

Nonetheless, the problem focusing on learning codebooks simultaneously at different network nodes has not been explored. The existing approaches, for instance \cite{Zhang2022Reinforcement,Zhang2023Online}, can hardly be applied in this context due to their reliance on the stationary interference patterns as well as their centralized learning paradigms. This motivates the development of fully decentralized codebook learning approaches that do not require information exchange between different agents, which is the focus of this paper.

\textbf{Contribution:}
Designing analog codebook patterns for mmWave MIMO system is an important research problem and can be found critical for various use cases in practical systems.
The online learning of these codebooks suffers from the potential non-stationary interference caused by the adjacent networks, which makes the codebook learning task rather challenging.
In this paper, we propose a fully decentralized multi-agent deep reinforcement learning based codebook learning framework that can efficiently adapt the codebook patterns at different networks without sharing their local information to each other.
This is done by including an averaging process of the power measurements for the reward generation to mitigate the non-stationarity effect. 
The justification of the averaging is also theoretically examined and is proven to be sufficient in an asymptotic sense.
The main contributions of this paper can be summarized as follows:
\begin{itemize}
  \item \textbf{Generalizing the beam codebook learning problem} to scenarios with multiple networks, where the operations of the adjacent base stations cause non-stationary interference to the target base station. We provide a new formulation for the codebook learning in such scenario. The core problem for this problem is identified.
  \item \textbf{Proposing a multi-agent beam codebook learning framework} that is able to operate in a fully decentralized manner and without any information exchange between different agents. To achieve that, a novel reward function based on measurement averaging is proposed for mitigating the non-stationarity effect.
  \item \textbf{Analyzing the performance of the power averaging method} in determining the interference suppression capability of a specific beam. We theoretically prove that the proposed estimator is a sufficient statistic for the underlying (i.e., not directly observable) quantity of interest in an asymptotic sense for mmWave large antenna array systems with limited multi-paths.
  \item \textbf{Extensively evaluating the proposed decentralized codebook learning solutions} using numerical simulations. This provides a comprehensive assessment of the capability of the proposed learning approaches in learning site-specific codebooks without information sharing between networks.
\end{itemize}

The simulation results show that the developed fully decentralized codebook learning solution can effectively adapt codebook beam patterns at different network nodes, without requiring any coordination and information exchange between the learning agents.
By using only local power measurement information, the proposed learning framework significantly relaxes the system requirements and reduces the overall deployment complexities.
This leads to the enhanced scalability of the future mmWave/THz networks, and efficiently manages the interference in application scenarios involving dense deployments.
The learned codebooks at different base stations demonstrate promising beam patterns in nulling the undesired interfering directions as well as in preventing the creation of interference to others.

\begin{figure*}[t]
	\centering
	\includegraphics[width=.85\textwidth]{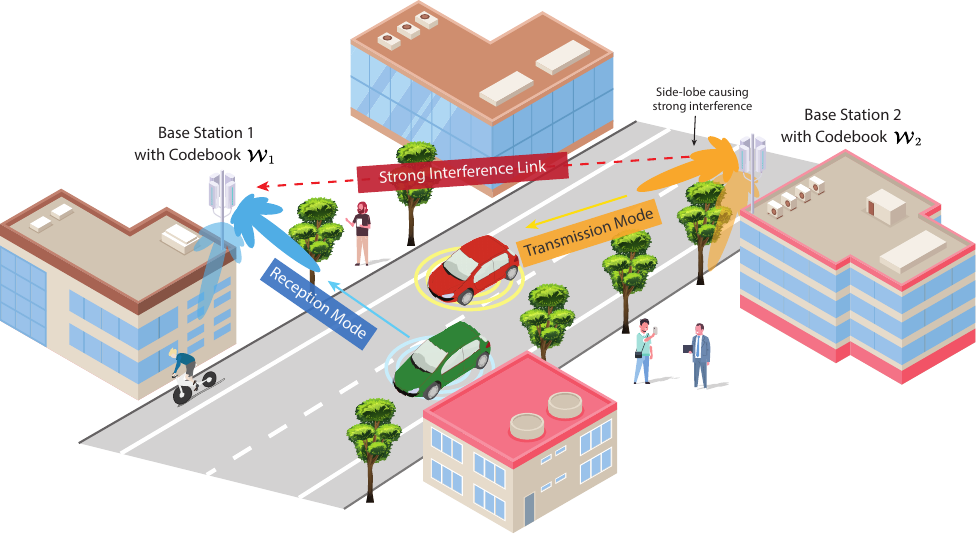}
    \caption{The considered scenario where there are multiple mmWave base stations operating at the same time and frequency to serve the surrounding users. Besides, there is no coordination among those base stations (such as user scheduling, power control, etc.) and no information sharing, which leads to unavoidable in-band interference that limits the system performance.}
	\label{fig:sys}
\end{figure*}

\section{System and Channel Models} \label{sec:System}

In this section, we introduce in detail the adopted system and channel models for the considered decentralized codebook learning problem.

\subsection{System Model}

We consider a communication system where there are $K$ mmWave MIMO base stations (BSs), each equipped with $M$ antennas and operating in time-division duplex (TDD) mode.
Each BS $k$ is assumed to serve single-antenna user equipments (UEs) from a candidate grid $\boldsymbol{\mathcal{H}}_k, \forall k\in\{1,2,\dots,K\}$.
Besides, we consider a practical scenario of asynchronous BS transmission. This could model the case when the time synchronization between the BSs of the same network is not perfect or the case when the BSs belong to different operators/networks that have no common transmission policy.
As a result, in-band interference may occur between the BSs, particularly those in close spatial proximity.
It is worth highlighting that such interference could be high, since the BSs always use much higher transmit power than the UEs, and they normally have direct LOS link between each other due to their high elevations, as illustrated in \fref{fig:sys}.

Without loss of generality, we assume that at a given time instant, the BS $k$ is \textbf{receiving} signal from UE $u_k$ (randomly selected from $\boldsymbol{\mathcal{H}}_k$, i.e., $u_k\in\{1,2,\dots, |\boldsymbol{\mathcal{H}}_k|\}$) and the other BSs ($q\ne k$)\update{\footnote{\update{In cases when not all the BSs (except the $k$-th one) are in the transmission mode, the summation in \eqref{rec-sig} is simply over those BSs that are transmitting signals. Moreover, we assume that the set of interfering BSs are always the same due to the fixed frame structure.}}} are \textbf{transmitting} signals to their served users $u_q\in \boldsymbol{\mathcal{H}}_q$.
Hence, the signal received at BS $k$ after combining can be expressed as
\begin{equation}\label{rec-sig}
  y_k = \mathbf{w}^H\mathbf{h}_{u_k}x_{u_k} + \sum_{q\ne k}\mathbf{w}^H\mathbf{H}_{qk}\mathbf{f}_qx_{u_q} + \mathbf{w}^H\mathbf{n}_k,
\end{equation}
where $\mathbf{h}_{u_k}\in\mathbb{C}^{M\times 1}$ is the channel vector between BS $k$ and UE $u_k$, $\mathbf{H}_{qk}\in\mathbb{C}^{M\times M}$ is the channel matrix between BS $k$ and BS $q$.
It is worth pointing out here that, for clarity, we subsume the factors such as path-loss and transmission power into the channels.
$x_{u_k}\in\mathbb{C}$ is the transmitted symbol from UE $u_k$ to BS $k$, and $x_{u_q}\in\mathbb{C}$ is the transmitted symbol from BS $q$ to UE $u_q$.
All transmitted symbols satisfy the average power constraint $\mathbb{E}\left[|x_{u_k}|^2\right]=P_x, \forall k\in\{1, 2, \dots, K\}$.
The vector ${\bf w}\in\mathbb{C}^{M\times 1}$ denotes the combining vector used by BS $k$, and ${\bf f}_q\in\mathbb{C}^{M\times 1}$ is the transmit beamforming vector used by BS $q$. ${\bf n}_k \sim \mathcal{CN}(0, \sigma^2{\bf I})$ is the receive noise vector at BS $k$ with $\sigma^2$ denoting the noise power.

Further, given the high cost and power consumption of the mixed-signal components, we consider a practical system where all the BSs have only one radio frequency (RF) chain
and employ analog-only beamforming/combining using a network of $r$-bit quantized phase shifters.
Moreover, to facilitate the system operation and respect the hardware constraints, mmWave MIMO systems typically rely on pre-defined beamforming/combining codebooks in serving their users. Let $\boldsymbol{\mathcal{W}}_k$ denote the beam codebook adopted by BS $k$ and assume that it contains $N$ beamforming/combining vectors, each of which takes the form
\begin{equation}\label{Analog}
  {\bf w} = \frac{1}{\sqrt{M}}\left[ e^{j\theta_1}, e^{j\theta_2}, \dots, e^{j\theta_M} \right]^T,
\end{equation}
where each phase shift $\theta_m$ is selected from a finite set $\boldsymbol{\Psi}$ with $2^r$ possible discrete values drawn uniformly from $(-\pi, \pi]$. The normalization factor $M^{-1/2}$ is to make sure the beamformer has unit power, i.e., $\|\mathbf{w}\|_2^2=1$.
Without loss of generality, we assume that BS $q, \forall q\ne k$ also has a codebook, denoted as $\boldsymbol{\mathcal{W}}_q$, that contains $N$ beams as well. The transmit beamforming vector $\mathbf{f}_q\in\boldsymbol{\mathcal{W}}_q$ used by BS $q$ can be expressed in a similar fashion as \eqref{Analog}.

Furthermore, the selection of the beams $\mathbf{w}$ and $\mathbf{f}_q, \forall q\ne k$ in \eqref{rec-sig} is based on the regular beam training procedure conducted in the practical mmWave MIMO system. To be more specific, for BS $k$, the beam is selected based on the following criterion\footnote{\highlight{The discussion about how to estimate the desired signal power under the presence of interference is provided in \sref{special}.}}
\begin{equation}\label{beam-training-w}
  \mathbf{w} = \mathop{\arg\max}_{\mathbf{w}\in\boldsymbol{\mathcal{W}}_k}|{\bf w}^H{\bf h}_{u_k}|^2, ~~ \forall u_k \in \boldsymbol{\mathcal{H}}_k,
\end{equation}
and for BS $q, \forall q\ne k$, it follows the similar process
\begin{equation}\label{beam-training-f}
  \mathbf{f}_q = \mathop{\arg\max}_{\mathbf{f}\in\boldsymbol{\mathcal{W}}_q}|{\bf f}^H\mathbf{h}_{u_q}|^2, ~~ \forall u_q \in \boldsymbol{\mathcal{H}}_q.
\end{equation}
%
As can be seen from \eqref{beam-training-w} and \eqref{beam-training-f}, the selection of beam made by each base station is solely based on the performance with respect to its own desired user, which is common in an asynchronous network with no coordinations.
\update{
It is worth emphasizing that we do not assume that a BS will use the same codebook for uplink reception and downlink transmission.
However, in cases where they do use the same codebook in uplink and downlink,
%
%
then, if BS $k$ is transmitting signal to the UE $u_k$, it will select a beamforming vector $\mathbf{w}$\footnote{We always discuss from BS $k$'s perspective, which can be any one of those $K$ base stations in the network. Therefore, without confusion, we denote the beam used by BS $k$ as $\mathbf{w}$, whether it is used for transmission or reception. Similarly, regardless of its current mode, the beam used by BS $q$ is denoted as $\mathbf{f}_q$.} from $\boldsymbol{\mathcal{W}}_k$. And correspondingly, when BS $q$ is receiving signal from the UE $u_q$, it will use a combining vector $\mathbf{f}_q$ from $\boldsymbol{\mathcal{W}}_q$.
The resulting received symbol at the BS $q$ after combining can be written similarly as \eqref{rec-sig}.
}

\subsection{Channel Model}

We adopt a narrowband geometric channel model for both the channels between BSs and their UEs, as well as the interference channels between any pair of base stations. For the channels between the BSs and their served UEs, they admit the same form, as given below
\begin{equation}\label{ch}
  {\bf h}_{u_k} = \sum_{\ell=1}^{L_{u_k}}\alpha_{u_k, \ell}{\bf a}(\phi_{u_k, \ell}, \vartheta_{u_k, \ell}), ~ \forall u_k \in \{1,2,\dots,|\boldsymbol{\mathcal{H}}_k|\}, \forall k \in \{1,2,\dots,K\},
\end{equation}
where we assume that the signal propagation between UE $u_k$ and BS $k$ consists of $L_{u_k}$ multi-paths. Each path $\ell$ has a complex gain $\alpha_{u_k, \ell}$, which subsumes the factors such as path-loss, transmission power, etc. The angles $\phi_{u_k, \ell}$ and $\vartheta_{u_k, \ell}$ represent the $\ell$-th path's azimuth and elevation angles of arrival (or departure) respectively,
%
%
and ${\bf a}(\phi_{u_k, \ell}, \vartheta_{u_k, \ell})$ is the array response vector of the considered BS to the signal with an angle of arrival (or departure) of $\phi_{u_k, \ell}$ and $\vartheta_{u_k, \ell}$ (with respect to the azimuth and elevation directions). It is also worth emphasizing that the array response vector ${\bf a}(\cdot,\cdot)$ might take different forms for different BSs, depending on their specific antenna array geometries as well as possible hardware impairments. We do not distinguish such difference in \eqref{ch} just for clarity.

For the interference channel matrix between any two BSs, for instance, BS $k$ and BS $q$, similarly, we assume that the signal propagation between these two BSs consists of $L_{qk}$ multi-paths. Then, the interference channel matrix $\mathbf{H}_{qk}$ can be expressed as
\begin{equation}\label{ch-matrix}
  \mathbf{H}_{qk} = \sum_{\ell=1}^{L_{qk}}\alpha_{qk, \ell}\mathbf{a}_\mathrm{r}(\phi_{qk, \ell}^\mathrm{r}, \vartheta_{qk, \ell}^\mathrm{r})\mathbf{a}_\mathrm{t}^H(\phi_{qk, \ell}^\mathrm{t}, \vartheta_{qk, \ell}^\mathrm{t}),
\end{equation}
where, similarly, $\alpha_{qk, \ell}$ denotes the complex path gain of the $\ell$-th path, which subsumes the factors such as path-loss, transmission power, etc. The angles $\phi_{qk, \ell}^\mathrm{r}, \vartheta_{qk, \ell}^\mathrm{r}$ represent the $\ell$-th path's azimuth and elevation angle of arrival at the receiver respectively, and $\phi_{qk, \ell}^\mathrm{t}, \vartheta_{qk, \ell}^\mathrm{t}$ represent the $\ell$-th path's azimuth and elevation angle of departure from the transmitter respectively.
Depending on the specific transmission/reception modes that the two base stations are currently in, $\mathbf{a}_\mathrm{r}$ and $\mathbf{a}_\mathrm{t}$ in \eqref{ch-matrix} take the corresponding forms of the array response vectors of BS $k$ and BS $q$'s, respectively.

\ifnum\x=0
{\clearpage}
\fi

\section{Problem Formulation} \label{sec:Prob}

In this paper, we investigate the problem of learning interference-aware beam codebook under the presence of non-stationary interference sources.
To be more specific, we consider the setting where the interfering BSs are switching their beams during the beam codebook measurement/learning process of the target BS.
For ease of exposition, we will focus on the case when only two BSs exist.
\update{
It is worth mentioning, however, that the formulated problem and developed solution are also applicable to the more general cases. This is because when more than two BSs exist, all the interfering BSs can be collectively treated as one interfering BS from the receivers' perspective.
}

Assume that at a given time instant, BS 1 is receiving signal from UE $u_1$ (randomly selected from $\boldsymbol{\mathcal{H}}_1$) and BS 2 is transmitting signal to UE $u_2$ (randomly selected from $\boldsymbol{\mathcal{H}}_2$). Hence, the signal received by BS 1 after combining can be expressed as
\begin{equation}\label{rec-sig-bs1}
  y_1 = \mathbf{w}^H\mathbf{h}_{u_1}x_{u_1} + \mathbf{w}^H\mathbf{H}\mathbf{f}x_{u_2} + \mathbf{w}^H\mathbf{n}_1.
\end{equation}
As can be seen, \eqref{rec-sig-bs1} is a special case of \eqref{rec-sig}. Moreover, since there are only two BSs, we ignore the subscript of the interference channel matrix in \eqref{rec-sig} and use $\mathbf{H}\in\mathbb{C}^{M\times M}$ to denote the channel between BS 1 and BS 2.
Based on \eqref{rec-sig-bs1}, the average achievable rate of UE $u_1$ can be written as
\begin{equation}\label{rate-bs1}
  R_1(\mathbf{h}_{u_1}) = \mathbb{E}_{\mathbf{f}\in\boldsymbol{\mathcal{W}}_2}\left[\log_2\left(1+\frac{|{\bf w}^H{\bf h}_{u_1}|^2P_x}{|\mathbf{w}^H\mathbf{H}\mathbf{f}|^2P_x + \sigma^2}\right)\right].
\end{equation}
It is worth noting that the expectation is over the transmit beamforming vector used by BS 2.
Similarly, when BS 2 is in the reception mode and BS 1 is in the transmission mode, the average achievable rate at BS 2 of UE $u_2$ is given by
\begin{equation}\label{rate-bs2}
  R_2(\mathbf{h}_{u_2}) = \mathbb{E}_{\mathbf{w}\in\boldsymbol{\mathcal{W}}_1}\left[\log_2\left(1+\frac{|{\bf f}^H{\bf h}_{u_2}|^2P_x}{|\mathbf{f}^H\mathbf{H}^H\mathbf{w}|^2P_x + \sigma^2}\right)\right].
\end{equation}
Note that $\mathbf{w}$ in \eqref{rate-bs1} and $\mathbf{f}$ in \eqref{rate-bs2} are determined via the beam training process, as described in \sref{sec:System}.
Based on \eqref{rate-bs1} and \eqref{rate-bs2}, we formulate the considered interference-aware beam codebook learning problem as
\begin{align}\label{prob-major}
  \boldsymbol{\mathcal{W}}_1^\star, \boldsymbol{\mathcal{W}}_2^\star =
  \mathop{\arg\max}_{\boldsymbol{\mathcal{W}}_1, \boldsymbol{\mathcal{W}}_2} & ~ \left(\frac{1}{|\boldsymbol{\mathcal{H}}_1|}\sum_{\mathbf{h}_{u_1}\in\boldsymbol{\mathcal{H}}_1}R_1(\mathbf{h}_{u_1}) +
  \frac{1}{|\boldsymbol{\mathcal{H}}_2|}\sum_{\mathbf{h}_{u_2}\in\boldsymbol{\mathcal{H}}_2}R_2(\mathbf{h}_{u_2})\right), \\
  \subto & ~ \arg\left([\mathbf{w}]_m\right) \in \boldsymbol{\Psi}, ~ \forall m=1,2,\dots,M, ~ \forall \mathbf{w}\in\boldsymbol{\mathcal{W}}_1, \label{const-1} \\
  & ~ \arg\left([\mathbf{f}]_m\right) \in \boldsymbol{\Psi}, ~ \forall m=1,2,\dots,M, ~ \forall \mathbf{f}\in\boldsymbol{\mathcal{W}}_2, \label{const-2} \\
  & ~ \left|[\mathbf{w}]_m\right| = \left|[\mathbf{f}]_m\right| = \frac{1}{\sqrt{M}}, ~ \forall m=1,2,\dots,M, ~ \forall \mathbf{w}\in\boldsymbol{\mathcal{W}}_1, \forall \mathbf{f}\in\boldsymbol{\mathcal{W}}_2,
\end{align}
where $\arg(\cdot)$ takes the argument of a complex number. As can be seen in \eqref{prob-major}, we cast the objective of the codebook design problem as maximizing the average achievable rate of the two BSs over their user grids.
The outcome of solving this problem is expected to be two beam codebooks that compromise between improving the desired signal power received from their own users and suppressing the undesired interference incurred from the other BS.

In addition to the difficulties encountered in the practical fully analog transceiver architecture, such as the non-convex and discrete phase shifter constraints \eqref{const-1} and \eqref{const-2}, and the unavailability of the channels, \eqref{prob-major} brings another two major challenges in its design process:
\textbf{First,} designing a codebook is associated with higher computational complexity compared to designing a single beam pattern.
\textbf{Second,} due to the non-cooperative nature of the asynchronous wireless network, there is \emph{no information exchange} between the BSs.
For example, a BS does not know which beam is currently used by the other BS. 
Such fully decentralized and independent operations of the different BSs result in the non-stationary interference behavior of the environment.
To address these challenges, we develop a decentralized deep reinforcement learning-based algorithm that is able to tackle the beam codebook learning problem while respecting all the aforementioned constraints, which will be described in detail in the next section.

\ifnum\x=0
{\clearpage}
\fi

\section{Decentralized Reinforcement Learning Solution} \label{sec:Sol-CB}

In this section, we describe in detail the proposed fully decentralized multi-agent deep reinforcement learning based interference-aware beam codebook learning approach.
We start by making the following observations on the beam codebook learning problem in \eqref{prob-major}.
\textbf{First}, due to the fully independent operation of the two (or multiple in general) BSs, the algorithm can be developed from any BS's perspective and the acquired solution should be applicable to other BS(s).
\textbf{Second}, the codebook learning problem can be decomposed into multiple \emph{independent} beam pattern learning sub-problems, which makes the complexity of the composite problem only linearly scale with the codebook size, as in the previous work \cite{Zhang2022Reinforcement}.
Based on these two important observations, we will show the core problem that resides in the original problem \eqref{prob-major} in the next subsection, which is learning interference-aware beam pattern under the presence of the non-cooperative and non-stationary interfering sources.
We then discuss how a BS can determine the interference suppression capability of its beams in \sref{subsec:sens}, which motivates the adoption of the proposed reward mechanism that will be described in \sref{special}.

\subsection{Beam Learning Under Non-Stationary Interference}

Based on the above observations, the core problem is identified, which will be serving as a sub-problem of the original problem \eqref{prob-major}.
\textbf{To reflect the first observation}, we focus on the case where BS 1 is receiving signals from UEs, and BS 2 is transmitting signals to the UEs which causes the interference to BS 1.
Moreover, \textbf{to reflect the second observation}, we adopt the power-based user clustering and assignment algorithm proposed in \cite{Zhang2022Reinforcement} to partition the user channels in $\boldsymbol{\mathcal{H}}_1$ into $N$ disjoint groups, i.e., $\boldsymbol{\mathcal{H}}_1 = \boldsymbol{\mathcal{H}}_1^{(1)} \cup \boldsymbol{\mathcal{H}}_1^{(2)} \cup \cdots \cup \boldsymbol{\mathcal{H}}_1^{(N)}$, where each user group is supposed to be served by one of the beams in the codebook $\boldsymbol{\mathcal{W}}_1$.
Thanks to the user partitioning mechanism, the original codebook learning problem is decomposed into $N$ independent beam learning sub-problems.
As a result, we only focus on the $n$-th beam in the codebook and cast the interference-aware beam design problem as
\begin{align}\label{prob-core}
 {\bf w}_n^\star = \argmax\limits_{{\bf w}} & \hspace{2pt}  \frac{1}{|\boldsymbol{\mathcal{H}}_1^{(n)}|} \sum_{\mathbf{h}_{u_1}\in\boldsymbol{\mathcal{H}}_1^{(n)}} \mathbb{E}_{\mathbf{f}\in\boldsymbol{\mathcal{W}}_2(t)}\left[ \log_2\left(1 + \frac{|{\bf w}^H{\bf h}_{u_1}|^2P_x}{|\mathbf{w}^H\mathbf{H}\mathbf{f}|^2P_x + \sigma^2} \right) \right], \\
 \subto  \hspace{2pt} &  [\mathbf{w}]_{m} = \frac{1}{\sqrt{M}}e^{j\theta_{m}}, ~ \forall m=1, ..., M, \label{pc1} \\
 & \theta_{m}\in\boldsymbol{\Psi}, ~ \forall m=1, ..., M, \label{pc2}
\end{align}
where $\boldsymbol{\mathcal{H}}_1^{(n)}$ is the $n$-th user cluster and is supposed to be served by using the $n$-th beam in the codebook.
We use the expectation in \eqref{prob-core} to characterize the non-stationary behavior of BS 2, i.e., it does not keep using the same beam.
Moreover, the codebook used by BS 2, i.e., $\boldsymbol{\mathcal{W}}_2(t)$, is not assumed to be fixed and it could change over time.
In other words, we \textbf{do not assume} that BS 2 always uses the same codebook (or suspends its own codebook update) during the course of BS 1's codebook learning process, as this is not realistic in an asynchronous network where the different BSs might belong to different operators and hence cannot control or communicate with each other.
It is worth mentioning, though, that the formulated problem does include the special case when $\boldsymbol{\mathcal{W}}_2$ is a fixed codebook, for instance, a beamsteering codebook. Therefore, the developed algorithm also applies therein.

The major difficulty that makes the problem \eqref{prob-core} quite challenging is the presence of the non-stationary interference transmitter.
To be more specific, due to the existence of the non-stationary interfering source (i.e., BS 2), the environment from BS 1's perspective is no longer stationary.
Such non-stationarity of the environment makes the existing reinforcement learning based codebook design approach, e.g., \cite{Zhang2022Reinforcement}, hard to converge, as the underlying Markov Decision Process (MDP) that the algorithm is built upon is no longer valid.
In other words, the agent could experience different (or even contradictive) reward feedback (i.e., the resulting Q values) in evaluating the same action.
To reduce such non-stationarity, it normally requires either a centralized control or a certain level of information sharing among agents (i.e., different BSs).
For instance, centralized training and decentralized execution framework proposed in \cite{Lowe2017} has been widely adopted as an effective approach in training multi-agent systems. This is achieved by leveraging a central critic under the assumption of availability of global information of all agents to avoid the non-stationary problem.
However, such centralized framework can hardly be applied in the considered system, where there is no central controller that jointly operates those BSs and there is no information sharing among BSs, as they might belong to different operators.
\textbf{
As a result, the problem considered in \eqref{prob-major} requires a fully decentralized approach such that each BS can evaluate its own behavior based on its partial view of the environment and is able to improve its policy accordingly.
}
\highlight{
To this end, we choose to design a \textbf{robust} reward mechanism, such that the quality of a BS's action can be accurately reflected, with lowest possible influence by the other BSs' behaviors.
}
Therefore, in the next subsection, we will present the core idea used in this paper to tackle the non-stationarity dilemma, which forms the vital step towards the finally proposed fully decentralized multi-agent reinforcement learning based approach.

\subsection{Estimating the Interference Suppression Performance} \label{subsec:sens}

In this subsection, we provide discussions on how to estimate the capability of a combining beam $\mathbf{w}$ in suppressing the interference, hence related to its signal-to-interference-plus-noise ratio (SINR) performance, in the presence of the non-stationary interference source.
The overall objective is to find an evaluation method such that each individual BS has a better understanding of the quality of its own beam in suppressing the interference signals, aiming to minimize the influence brought by the other interfering BSs as much as possible.
To this end, we try to investigate the potential relationship between the interference power and the combining vector used by the BS.
Specifically, by substituting the interference channel model \eqref{ch-matrix} into the objective function of \eqref{prob-core}, the interference term at the denominator can be explicitly written as
\begin{align}\label{interf-term}
  \left|\mathbf{w}^H\mathbf{H}\mathbf{f}\right|^2 = & \left|\mathbf{w}^H \left( \sum_{\ell=1}^{L}\alpha_\ell\mathbf{a}_\mathrm{r}(\phi_{\mathrm{r}, \ell}, \vartheta_{\mathrm{r}, \ell})\mathbf{a}_\mathrm{t}^H(\phi_{\mathrm{t}, \ell}, \vartheta_{\mathrm{t}, \ell}) \right) \mathbf{f}\right|^2, \\
  = & \left|\sum_{\ell=1}^{L}\alpha_\ell \cdot \mathbf{w}^H\mathbf{a}_\mathrm{r}(\phi_{\mathrm{r}, \ell}, \vartheta_{\mathrm{r}, \ell}) \cdot \mathbf{a}_\mathrm{t}^H(\phi_{\mathrm{t}, \ell}, \vartheta_{\mathrm{t}, \ell})\mathbf{f}\right|^2, \label{simp} \\
  = & \left|\sum_{\ell=1}^{L}\alpha_\ell \cdot \beta_\ell \cdot z_\ell\right|^2, \\
  = & \left|\boldsymbol{\xi}^T\mathbf{z}\right|^2, \label{vecnorm}
\end{align}
where we remove the subscript for indexing different pairs of BSs in \eqref{ch-matrix} as we only consider two BSs \highlight{and we ignore the average transmitted symbol power $P_x$, which is a constant, in the following discussions.}
Furthermore, we simplify \eqref{simp} by first defining $\beta_\ell = \mathbf{w}^H\mathbf{a}_\mathrm{r}(\phi_{\mathrm{r}, \ell}, \vartheta_{\mathrm{r}, \ell})$ and $z_\ell = \mathbf{a}_\mathrm{t}^H(\phi_{\mathrm{t}, \ell}, \vartheta_{\mathrm{t}, \ell})\mathbf{f}$.
Then, we use their vector form, i.e., $\boldsymbol{\xi} = [\alpha_1\beta_1, \dots, \alpha_L\beta_L]^T\in\mathbb{C}^{L\times 1}$ and $\mathbf{z} = [z_1, \dots, z_L]^T\in\mathbb{C}^{L\times 1}$ in \eqref{vecnorm} to further simplify the expression of the interference power.
\highlight{
It is beneficial to view these two $L\times 1$ vectors, i.e., $\boldsymbol{\xi}$ and $\mathbf{z}$, as in the ``multi-path'' space.
}
It is also worth emphasizing that we have the implicit inequalities $0\le|\beta_\ell|\le1$ and $0\le|z_\ell|\le1$ and we subsume the factors such as path-loss and transmission power into the path gain $\alpha_\ell$.

\highlight{
As can be seen, $\beta_\ell$ essentially characterizes how effective a combining vector is in suppressing the interference signal coming from the $\ell$-th impinging direction. Weighted by the path gain $\alpha_\ell$, the vector $\boldsymbol{\xi}$ finally determines the overall performance of the given combining vector in all interference directions.
}
Therefore, we cast the interference suppression evaluation task as a hypothesis testing problem as follows.
Specifically, assume that there are two combining vectors, i.e., $\mathbf{w}$ and $\mathbf{w}^\prime$, that correspond to $\boldsymbol{\xi}$ and $\boldsymbol{\xi}^\prime$ respectively. We aim to derive a decision rule for the following hypothesis testing problem
\begin{align}
  H_0 \hspace{1pt}:& \hspace{5pt} \|\boldsymbol{\xi}\|_2^2 < \|\boldsymbol{\xi}^\prime\|_2^2, \label{hypo-0} \\
  H_1 \hspace{1pt}:& \hspace{5pt} \|\boldsymbol{\xi}\|_2^2 \ge \|\boldsymbol{\xi}^\prime\|_2^2, \label{hypo-1}
\end{align}
where, by definition
\begin{equation}\label{xi-norm}
  \|\boldsymbol{\xi}\|_2^2=\sum_{\ell}|\alpha_\ell|^2\left|\mathbf{w}^H\mathbf{a}_\mathrm{r}(\phi_{\mathrm{r}, \ell}, \vartheta_{\mathrm{r}, \ell})\right|^2.
\end{equation}
In short, the hypothesis testing problem \eqref{hypo-0} and \eqref{hypo-1} determines the belief on which combining vector is better, in terms of the effectiveness in suppressing the signals received from the unknown interfering directions.
\highlight{
The available sample data to such hypothesis testing problem could possibly be based on the two sets of measurements $\mathcal{I}(\mathbf{w})=\{|\boldsymbol{\xi}^T\mathbf{z}_1|^2, |\boldsymbol{\xi}^T\mathbf{z}_2|^2, \dots, |\boldsymbol{\xi}^T\mathbf{z}_K|^2\}$ and $\mathcal{I}(\mathbf{w}^\prime)=\{|{\boldsymbol{\xi}^\prime}^T\mathbf{z}_1^\prime|^2, |{\boldsymbol{\xi}^\prime}^T\mathbf{z}_2^\prime|^2, \dots, |{\boldsymbol{\xi}^\prime}^T\mathbf{z}_K^\prime|^2\}$, the details of which will be discussed in \sref{special}.
}

In order to find a meaningful decision rule for the hypothesis testing problem \eqref{hypo-0} and \eqref{hypo-1}, we examine the property of the interference power, i.e., $\left|\mathbf{w}^H\mathbf{H}\mathbf{f}\right|^2$, as it contains related information regarding the two measurement sets $\mathcal{I}(\mathbf{w})$ and $\mathcal{I}(\mathbf{w}^\prime)$.
Moreover, because BS 1 does not have any knowledge about the transmit beamforming vector (i.e., $\mathbf{f}$) used by BS 2, as well as the pattern by which BS 2 changes its beam, $\mathbf{f}$ is reasonably to be modeled as a random vector from BS 1's perspective. This makes $\mathbf{z}$ in \eqref{vecnorm} a statistical quantity.
Therefore, for any given receive combining vector $\mathbf{w}$ (which means that $\boldsymbol{\xi}$ is fixed), we inspect the expectation of the interference power, with respect to the unknown transmit beamforming vector of BS 2, i.e., $\mathbf{f}$, that is
\begin{equation}\label{var-quad}
  \mathbb{E}\left[\left|\mathbf{w}^H\mathbf{H}\mathbf{f}\right|^2\right]
  = \mathbb{E}\left[\left|\boldsymbol{\xi}^T\mathbf{z}\right|^2\right]
  = \mathbb{E}\left[\mathbf{z}^T\boldsymbol{\xi}\boldsymbol{\xi}^H\mathbf{z}^*\right]
  = \mathbb{E}\left[\mathbf{z}^H\boldsymbol{\xi}^*\boldsymbol{\xi}^T\mathbf{z}\right]
  = \mathbb{E}\left[\mathbf{z}^H\tilde{\boldsymbol{\xi}}\tilde{\boldsymbol{\xi}}^H\mathbf{z}\right],
\end{equation}
where $\tilde{\boldsymbol{\xi}}=\boldsymbol{\xi}^*$, and the information of $\mathbf{f}$ is partially contained in $\mathbf{z}$.
\highlight{
Given that $\|\tilde{\boldsymbol{\xi}}\|_2^2 = \|\boldsymbol{\xi}\|_2^2$, any decision rule derived based on $\tilde{\boldsymbol{\xi}}$ is the same as that of $\boldsymbol{\xi}$.
}
Moreover, as can be seen from \eqref{var-quad}, the interference power admits a quadratic form of $\mathbf{z}$.
To leverage such structure, we denote $\mathbb{E}[\mathbf{z}]=\mathbf{0}$ and $\mathrm{Var}[\mathbf{z}]=\boldsymbol{\Sigma}$.
The validity of $\mathbb{E}[\mathbf{z}]=\mathbf{0}$ can be examined by leveraging the definition of $\mathbf{z}$, where
\begin{equation}
  \mathbb{E}[z_\ell] = \mathbb{E}[\mathbf{a}_\mathrm{t}^H(\phi_{\mathrm{t}, \ell}, \vartheta_{\mathrm{t}, \ell})\mathbf{f}] = \mathbf{a}_\mathrm{t}^H(\phi_{\mathrm{t}, \ell}, \vartheta_{\mathrm{t}, \ell})\mathbb{E}[\mathbf{f}]=0, ~ \forall \ell.
\end{equation}
By further denoting $\mathbf{A}=\tilde{\boldsymbol{\xi}}\tilde{\boldsymbol{\xi}}^H$ and by noticing that $\mathbf{A}=\mathbf{A}^H$, \eqref{var-quad} has the following concise expression \cite{papoulis02}
\begin{equation}\label{varterm}
  \mathbb{E}\left[\mathbf{z}^H\mathbf{A}\mathbf{z}\right] = \hspace{1pt} \mathrm{Tr}\left(\mathbf{A}\boldsymbol{\Sigma}\right)
  = \hspace{1pt} \mathrm{Tr}\left(\tilde{\boldsymbol{\xi}}\tilde{\boldsymbol{\xi}}^H\boldsymbol{\Sigma}\right)
  = \hspace{1pt} \mathrm{Tr}\left(\tilde{\boldsymbol{\xi}}^H\boldsymbol{\Sigma}\tilde{\boldsymbol{\xi}}\right)
  = \hspace{1pt} \tilde{\boldsymbol{\xi}}^H\boldsymbol{\Sigma}\tilde{\boldsymbol{\xi}}.
\end{equation}
Interestingly, \eqref{varterm} implies a potential decision rule for the hypothesis testing problem \eqref{hypo-0} and \eqref{hypo-1}, where, intuitively, a smaller $\tilde{\boldsymbol{\xi}}$ (in terms of vector $2$-norm) leads to smaller expectation of the interference power.
Formally, we provide the following proposition.

\begin{prop} \label{prop-1}
  Assume that $\mathrm{Var}[\mathbf{z}]=\boldsymbol{\Sigma}\succ 0$, and there are $\tilde{\boldsymbol{\xi}}$ and $\tilde{\boldsymbol{\xi}^\prime}$, such that
  \begin{equation}\label{condition}
    \|\tilde{\boldsymbol{\xi}}\|_2^2 < \frac{\lambda_L(\mathbf{I} + \boldsymbol{\Pi})}{1+\left\|\boldsymbol{\Pi}\right\|_2}\|\tilde{\boldsymbol{\xi}^\prime}\|_2^2,
  \end{equation}
  where $\lambda_L(\cdot)$ denotes the $L$-th largest eigenvalue of a positive definite matrix, and $\boldsymbol{\Pi}\in\mathbb{C}^{L\times L}$ is defined as
  \begin{equation}\label{Pi}
                                   \left[
                                     \begin{array}{cccc}
                                       0 & \mathbf{a}_\mathrm{t}(\phi_{\mathrm{t}, 1}, \vartheta_{\mathrm{t}, 1})^H\mathbf{a}_\mathrm{t}(\phi_{\mathrm{t}, 2}, \vartheta_{\mathrm{t}, 2}) & \cdots & \mathbf{a}_\mathrm{t}(\phi_{\mathrm{t}, 1}, \vartheta_{\mathrm{t}, 1})^H\mathbf{a}_\mathrm{t}(\phi_{\mathrm{t}, L}, \vartheta_{\mathrm{t}, L}) \\
                                       \mathbf{a}_\mathrm{t}(\phi_{\mathrm{t}, 2}, \vartheta_{\mathrm{t}, 2})^H\mathbf{a}_\mathrm{t}(\phi_{\mathrm{t}, 1}, \vartheta_{\mathrm{t}, 1}) & 0 & \cdots & \mathbf{a}_\mathrm{t}(\phi_{\mathrm{t}, 2}, \vartheta_{\mathrm{t}, 2})^H\mathbf{a}_\mathrm{t}(\phi_{\mathrm{t}, L}, \vartheta_{\mathrm{t}, L}) \\
                                       \vdots & \vdots & \ddots & \vdots \\
                                       \mathbf{a}_\mathrm{t}(\phi_{\mathrm{t}, L}, \vartheta_{\mathrm{t}, L})^H\mathbf{a}_\mathrm{t}(\phi_{\mathrm{t}, 1}, \vartheta_{\mathrm{t}, 1}) & \mathbf{a}_\mathrm{t}(\phi_{\mathrm{t}, L}, \vartheta_{\mathrm{t}, L})^H\mathbf{a}_\mathrm{t}(\phi_{\mathrm{t}, 2}, \vartheta_{\mathrm{t}, 2}) & \cdots & 0 \\
                                     \end{array}
                                   \right].\notag
  \end{equation}
  Then $\mathbb{E}\left[\mathbf{z}^H\mathbf{A}\mathbf{z}\right]<\mathbb{E}\left[\mathbf{z}^H\mathbf{A}^\prime\mathbf{z}\right]$, with $\mathbf{A}=\tilde{\boldsymbol{\xi}}\tilde{\boldsymbol{\xi}}^H$ and $\mathbf{A}^\prime=\tilde{\boldsymbol{\xi}^\prime}\tilde{\boldsymbol{\xi}^\prime}^H$.
\end{prop}

\begin{proof}
  See \apref{append-a}.
\end{proof}

\pref{prop-1} provides a \textbf{necessary condition} of using \eqref{varterm}, i.e., the expectation of the interference power, as a measure to \emph{estimate} the performance of a beam in suppressing the interference.
Inspired by this, we adopt the following decision rule
\begin{align}
  \text{Decide}~H_0 \hspace{1pt}:& \hspace{5pt} \text{If}~\mathbb{E}\left[\left|\mathbf{w}^H\mathbf{H}\mathbf{f}\right|^2\right] < \mathbb{E}\left[\left|{\mathbf{w}^\prime}^H\mathbf{H}\mathbf{f}\right|^2\right], \label{dec-0} \\
  \text{Decide}~H_1 \hspace{1pt}:& \hspace{5pt} \text{If}~\mathbb{E}\left[\left|\mathbf{w}^H\mathbf{H}\mathbf{f}\right|^2\right] \ge \mathbb{E}\left[\left|{\mathbf{w}^\prime}^H\mathbf{H}\mathbf{f}\right|^2\right], \label{dec-1}
\end{align}
where the expectation in \eqref{dec-0} and \eqref{dec-1} is approximated by using the average value of the interference power measurements in practical scenarios, the details of which will be given in \sref{special}.
The estimated interference suppression capability will then be used in designing the reward mechanism that reduces the
the effect of the non-stationarity for the learning purposes.
Furthermore, there are several interesting observations worth mentioning based on \eqref{condition}.

\noindent\textbf{Remark 1}: If we define $\eta = \frac{\lambda_L(\mathbf{I} + \boldsymbol{\Pi})}{1+\left\|\boldsymbol{\Pi}\right\|_2}$ in \eqref{condition}, which satisfies $0\le \eta \le 1$, intuitively, it controls the \emph{resolution} of the proposed criterion that judges the quality of the current $\tilde{\boldsymbol{\xi}}$ compared to the previous one $\tilde{\boldsymbol{\xi}^\prime}$.

\noindent\textbf{Remark 2}: A special case of \eqref{condition} is when $L=1$, i.e., there is only one path (for example, a direct LOS path) between two BSs\footnote{\highlight{Despite being a special case, it is one of the most commonly encountered scenarios in practice. This is because the strongest interference in high-frequency bands normally comes from the direct LOS link, which has much higher power than the other NLOS links.}}. In that case, we have $\left\|\boldsymbol{\Pi}\right\|_2=0$, $\lambda_L(\mathbf{I} + \boldsymbol{\Pi})=1$, and hence $\eta=1$. This suggests that as long as $\|\tilde{\boldsymbol{\xi}}\|_2^2 < \|\tilde{\boldsymbol{\xi}^\prime}\|_2^2$, we have $\mathbb{E}\left[\mathbf{z}^H\mathbf{A}\mathbf{z}\right]<\mathbb{E}\left[\mathbf{z}^H\mathbf{A}^\prime\mathbf{z}\right]$.

To gain some insight from \eqref{condition}, \highlight{we assume that $\left\|\boldsymbol{\Pi}\right\|_2 < 1$ holds. We will soon show that this holds with high probability when the number of antennas is much larger than the number of multi-paths.} As a direct result from \eqref{weyl}, we have
\begin{equation}
  0 < \left(1 - \left\|\boldsymbol{\Pi}\right\|_2\right) \le \lambda_L(\mathbf{I} + \boldsymbol{\Pi}).
\end{equation}
Therefore, \eqref{condition} implies that
\begin{equation}\label{condition-2}
  \|\tilde{\boldsymbol{\xi}}\|_2^2 < \frac{1-\left\|\boldsymbol{\Pi}\right\|_2}{1+\left\|\boldsymbol{\Pi}\right\|_2}\|\tilde{\boldsymbol{\xi}^\prime}\|_2^2 = \eta^\prime\|\tilde{\boldsymbol{\xi}^\prime}\|_2^2 \le \eta\|\tilde{\boldsymbol{\xi}^\prime}\|_2^2,
\end{equation}
where we define $\eta^\prime = \frac{1-\left\|\boldsymbol{\Pi}\right\|_2}{1+\left\|\boldsymbol{\Pi}\right\|_2}$.
This suggests that the resolution of the proposed decision rule is closely related to $\left\|\boldsymbol{\Pi}\right\|_2$.
It turns out that the assumption $\left\|\boldsymbol{\Pi}\right\|_2 < 1$ holds \highlight{with high probability}, especially when $M\gg L$.
Moreover, we have the following asymptotic property of $\left\|\boldsymbol{\Pi}\right\|_2$.

\begin{prop} \label{prop-2}
Assume that $\phi_{\mathrm{t}, \ell}, \vartheta_{\mathrm{t}, \ell}, \forall \ell$ are independently and uniformly sampled from $[0, \pi]$, then, for $\forall \epsilon>0$, we have
\begin{equation}\label{mMIMO}
  \lim_{\frac{M}{L}\rightarrow +\infty}\mathbb{P}\left[\left|1-\frac{1-\left\|\boldsymbol{\Pi}\right\|_2}{1+\left\|\boldsymbol{\Pi}\right\|_2}\right| > \epsilon\right] = 0.
\end{equation}
\end{prop}


\begin{proof}
  See \apref{append-b}.
\end{proof}

It is worth mentioning that $M\gg L$ is normally a valid assumption in the massive MIMO system at high frequency regime, where the channel covariance features high dimensionality yet low rank structure \cite{Alkhateeb2014}.
Interestingly, based on \pref{prop-2}, a \textbf{sufficient condition} of adopting \eqref{varterm} to estimate the interference suppression performance of a beam can be derived, asymptotically. This is summarized in the following corollary.

\begin{corollary}
If $0<\mathbb{E}\left[\mathbf{z}^H\mathbf{A}\mathbf{z}\right]<\mathbb{E}\left[\mathbf{z}^H\mathbf{A}^\prime\mathbf{z}\right]$, then $\|\tilde{\boldsymbol{\xi}}\|_2^2 < \|\tilde{\boldsymbol{\xi}^\prime}\|_2^2$, when $\frac{M}{L}\rightarrow +\infty$.
\end{corollary}

\begin{proof}

According to \eqref{sigma-decomp}, we have
\begin{equation}
  \mathbb{E}\left[\mathbf{z}^H\mathbf{A}\mathbf{z}\right] = \tilde{\boldsymbol{\xi}}^H\boldsymbol{\Sigma}\tilde{\boldsymbol{\xi}} = \frac{1}{M}\|\tilde{\boldsymbol{\xi}}\|_2^2 + \frac{1}{M}\tilde{\boldsymbol{\xi}}^H\boldsymbol{\Pi}\tilde{\boldsymbol{\xi}},
\end{equation}
which suggests that
\begin{equation}\label{eq:ratio}
  \frac{\mathbb{E}\left[\mathbf{z}^H\mathbf{A}\mathbf{z}\right]}{\mathbb{E}\left[\mathbf{z}^H\mathbf{A}^\prime\mathbf{z}\right]}
  =\frac{\|\tilde{\boldsymbol{\xi}}\|_2^2 + \tilde{\boldsymbol{\xi}}^H\boldsymbol{\Pi}\tilde{\boldsymbol{\xi}}}
  {\|\tilde{\boldsymbol{\xi}^\prime}\|_2^2 + \tilde{\boldsymbol{\xi}^\prime}^H\boldsymbol{\Pi}\tilde{\boldsymbol{\xi}^\prime}}.
\end{equation}
Moreover, according to the trace inequality \cite{Mirsky1975}, the quadratic terms in \eqref{eq:ratio} are bounded as follows
\begin{equation}\label{eq:von}
  -\left\|\boldsymbol{\Pi}\right\|_2\|\tilde{\boldsymbol{\xi}}\|_2^2 \le \tilde{\boldsymbol{\xi}}^H\boldsymbol{\Pi}\tilde{\boldsymbol{\xi}} \le \left\|\boldsymbol{\Pi}\right\|_2\|\tilde{\boldsymbol{\xi}}\|_2^2,
\end{equation}
which holds similarly for $\tilde{\boldsymbol{\xi}^\prime}^H\boldsymbol{\Pi}\tilde{\boldsymbol{\xi}^\prime}$.
In addition, based on \eqref{ortho-norm} and from \eqref{eq:von}, it implies that
\begin{equation}
  \lim_{\frac{M}{L}\rightarrow +\infty} \tilde{\boldsymbol{\xi}}^H\boldsymbol{\Pi}\tilde{\boldsymbol{\xi}} = 0.
\end{equation}
Therefore, we have
\begin{equation}
  \lim_{\frac{M}{L}\rightarrow +\infty} \frac{\mathbb{E}\left[\mathbf{z}^H\mathbf{A}\mathbf{z}\right]}{\mathbb{E}\left[\mathbf{z}^H\mathbf{A}^\prime\mathbf{z}\right]} = \frac{\|\tilde{\boldsymbol{\xi}}\|_2^2}{\|\tilde{\boldsymbol{\xi}^\prime}\|_2^2},
\end{equation}
which concludes the proof.

\end{proof}

\ifnum\x=0
{\clearpage}
\fi

\subsection{Determining the Reward} \label{subsec:sol}

In this subsection, we describe the proposed fully decentralized reinforcement learning-based interference-aware beam codebook learning approach.
It is worth pointing out that as we decompose the original codebook learning problem \eqref{prob-major} into multiple independent beam pattern learning sub-problems \eqref{prob-core}, we focus on the design of the individual beam in the codebook.
Moreover, we adopt the similar reinforcement learning formulation as presented in the previous work \cite{Zhang2022Reinforcement}.
Specifically, the definitions of the state and the action are the same, i.e., they are both specified as the phase configurations of the system.
We also adopt a binary reward signal, i.e., the reward $r_t$ takes values from $\{+1, -1\}$.
\highlight{
However, to address the new challenge, i.e., the non-stationarity, a different reward generation mechanism is introduced, which is the focus of this subsection.
}

To develop beam patterns that are aware of the interference, the exact reward value assigned for a certain designed beam should depend on its joint performance on maximizing the desired signal and on suppressing the interference signal, when compared to the previous beam.
Moreover, due to the non-stationarity, the receiver needs to acquire multiple measurements in order to have a relatively accurate evaluation, as discussed in the previous subsection.
This is in stark contrast to the case when the interference sources are stationary, which also includes the special case when there is no interference.
Formally, the beamforming gain performance of the current beam $\mathbf{w}_t$ on the target user cluster can be obtained by averaging the power measurements of all the users within the cluster
\begin{equation}\label{avg-gain}
  g(\mathbf{w}_t) = \frac{1}{|\boldsymbol{\mathcal{H}}_1^{(n)}|}\sum_{\mathbf{h}_{u_1}\in\boldsymbol{\mathcal{H}}_1^{(n)}}\left|\mathbf{w}_t^H\mathbf{h}_{u_1}\right|^2.
\end{equation}
And the expected interference power, i.e., $\mathbb{E}\left[\left|\mathbf{w}_t^H\mathbf{H}\mathbf{f}\right|^2\right]$, is used to characterize the interference suppression capability of the same beam.
Based on these two quantities, the adopted reward generation mechanism can then be expressed as follows
\begin{equation}\label{reward}
  r_t = \left\{
     \begin{array}{l}
       +1, ~\mathrm{if}~\mathbb{E}\left[\left|\mathbf{w}_t^H\mathbf{H}\mathbf{f}\right|^2\right]<\mathbb{E}\left[\left|\mathbf{w}_{t-1}^H\mathbf{H}\mathbf{f}\right|^2\right]~\mathrm{\textbf{and}}~g(\mathbf{w}_t)>g(\mathbf{w}_{t-1}), \\
       -1, ~\mathrm{otherwise}. \\
     \end{array}
  \right.
\end{equation}
It is worth pointing out that the reward generation mechanism is not unique.
For instance, one could use the ratio
\begin{equation}\label{ratio-sir}
  \zeta(\mathbf{w}_t) = \frac{g(\mathbf{w}_t)}{\mathbb{E}\left[\left|\mathbf{w}_t^H\mathbf{H}\mathbf{f}\right|^2\right]}
\end{equation}
to represent the performance of the beam $\mathbf{w}_t$ and generate the reward based on it, that is
\begin{equation}\label{reward-sir}
  r_t = \left\{
     \begin{array}{l}
       +1, ~\mathrm{if}~\zeta(\mathbf{w}_t) > \zeta(\mathbf{w}_{t-1}), \\
       -1, ~\mathrm{otherwise}. \\
     \end{array}
  \right.
\end{equation}
We choose \eqref{reward} instead of \eqref{reward-sir} mainly because of its stability.
As can be seen, the criterion for generating a positive reward is more strict in \eqref{reward}.
In other words, empirical evidence indicates that using \eqref{reward-sir} is more likely to have the agent stuck at a local region where the SIR is high but the beamforming gain of the target user is not enough.
\highlight{
Next, we provide more discussions on how to acquire $g(\mathbf{w}_t)$ and $\mathbb{E}\left[\left|\mathbf{w}_t^H\mathbf{H}\mathbf{f}\right|^2\right]$ in a practical setting.
}

\subsection{Practical Operations} \label{special}

\highlight{
In this subsection, we discuss how to estimate $g(\mathbf{w}_t)$ and $\mathbb{E}\left[\left|\mathbf{w}_t^H\mathbf{H}\mathbf{f}\right|^2\right]$ that are used by the proposed beam learning algorithm.
Before delving into the details, it is important to mention that the proposed measurement method is developed based on the general case that the receive BS has no information about the interferers' behavior.
In other words, in cases where the receive BS has some prior knowledge about the interferers,
such as the transmission status or the codebook pattern, there might be other measurement solutions that are customized for the given scenarios and hence more efficient.
However, we do not investigate such possibilities in this paper.
}

Upon forming a new beam $\mathbf{w}_t$ at time instant $t$, we assume that BS 1 will use this beam to first measure multiple times of the interference strength (for example, $P$ times), which can be done by muting its target UEs.
The set of acquired interference power measurements can be denoted as\footnote{\highlight{Here, for clarity, we ignore the noise term, as the interference-to-noise ratio (INR) is assumed to be high and hence the interference power is dominant in those measurements.}}
\begin{equation}\label{int-seq}
  \mathcal{I}(\mathbf{w}_t) = \left\{|\mathbf{w}_t^H\mathbf{H}\mathbf{f}_1|^2, |\mathbf{w}_t^H\mathbf{H}\mathbf{f}_2|^2, \dots, |\mathbf{w}_t^H\mathbf{H}\mathbf{f}_P|^2\right\}.
\end{equation}
After that, the target UE starts transmitting reference signals, and the BS 1 will use the same beam to measure another $Q$ times of the signal plus interference strength, denoted as
\begin{equation}\label{sig-int-seq}
  \mathcal{SI}(\mathbf{w}_t) = \left\{|\mathbf{w}_t^H\mathbf{h}_{u_1}|^2+|\mathbf{w}_t^H\mathbf{H}\mathbf{f}_1^\prime|^2, |\mathbf{w}_t^H\mathbf{h}_{u_1}|^2+|\mathbf{w}_t^H\mathbf{H}\mathbf{f}_2^\prime|^2, \dots, |\mathbf{w}_t^H\mathbf{h}_{u_1}|^2+|\mathbf{w}_t^H\mathbf{H}\mathbf{f}_Q^\prime|^2\right\}.
\end{equation}
Based on these two sets of power measurements, the beamforming gain achieved at UE $u_1$ of the current beam $\mathbf{w}_t$ can be estimated as
\begin{equation}
  |\mathbf{w}_t^H{\bf h}_{u_1}|^2 \approx \mathrm{mean}\left(\mathcal{SI}(\mathbf{w}_t)\right) - \mathrm{mean}\left(\mathcal{I}(\mathbf{w}_t)\right),
\end{equation}
where $\mathrm{mean}(\cdot)$ calculates the average of the numbers within a set.
And $g(\mathbf{w}_t)$ can then be determined through \eqref{avg-gain}.
\highlight{
It is worth pointing out that due to the similarity of the user channels within one group, the summation in \eqref{avg-gain} can normally just be performed over a subset of $\boldsymbol{\mathcal{H}}_1^{(n)}$. Moreover, such subset does not need to be fixed, i.e., it can be randomly sampled from $\boldsymbol{\mathcal{H}}_1^{(n)}$ each time when $g_t(\mathbf{w}_t)$ needs to be evaluated.
}
By leveraging \eqref{int-seq} and \eqref{sig-int-seq}, the expectation of the interference power can be approximated as
\begin{equation}\label{var-approx}
  \mathbb{E}\left[\left|\mathbf{w}_t^H\mathbf{H}\mathbf{f}\right|^2\right] \approx \frac{P\cdot\mathrm{mean}\left(\mathcal{I}(\mathbf{w}_t)\right) + Q\cdot\mathrm{mean}\left(\mathcal{SI}(\mathbf{w}_t)\right)-Q\cdot|\mathbf{w}_t^H{\bf h}_{u_1}|^2}{P+Q}.
\end{equation}
As can be seen, $g(\mathbf{w}_t)$ and $\mathbb{E}\left[\left|\mathbf{w}_t^H\mathbf{H}\mathbf{f}\right|^2\right]$, and hence the reward signal, can be acquired in such a way that \textbf{does not} rely on any explicit channel state information of both the target UEs and the interfering BS, and \textbf{does not} require any coordinations and information sharing between different BSs.
This makes the developed solution fully compatible with the considered asynchronous system and be able to operate in a fully decentralized manner.

\noindent\textbf{Remark}:
An important problem is how to reduce the measurement overhead in order to improve the overall learning efficiency. One possible way is through data sharing.
For instance, a closer look at the interference measurements, i.e., \eqref{int-seq}, indicates that sharing information among different beam learning agents (within one system) is possible.
\highlight{
To be more specific, the interference measurement data of a certain beam collected by the $n$-th beam learning agent, denoted as $\mathcal{I}^{(n)}(\mathbf{w})$, can be used by the other different beam learning agents for evaluating the interference suppression performance of the same beam, when encountered during the course of beam learning.
}
To achieve such purpose, for instance, the system can maintain a shared interference measurement dataset for each phase configuration of the antenna array, which can be modified (i.e., adding new measurement data) and accessed by all the beam learning agents.
This has the potential of reducing the beam learning overhead and providing more accurate estimation of the interference suppression capability of a beam.

\ifnum\x=0
{\clearpage}
\fi

\section{Simulation Results}

\begin{table}[t]
\caption{Hyper-parameters for channel generation}
\centering
\begin{tabular}{ c | c }
\hline\hline
	Parameter & Value \\
	\hline
    Name of scenario & O1\textunderscore60 \\
    Active BSs & 3 and 4 \\
    Active users & 800 to 1200 \\
    Number of antennas (x, y, z)  & (1, 16, 1) \\
    System bandwidth & 0.05 GHz \\
    Antenna spacing & 0.5 \\
    Number of OFDM sub-carriers & 512 \\
    OFDM sampling factor & 1 \\
    OFDM limit & 1 \\
    \hline\hline
\end{tabular}
\label{param}
\end{table}

\highlight{
In this section, we evaluate the performance of the proposed fully-decentralized interference-aware beam codebook learning solution.
We first introduce the simulation setup in \sref{sub:SS}.
Then, the details of the evaluation method is provided in \sref{sub:EVM}.
Finally, we present the simulation results in \sref{sub:NR-CB}.
}

\subsection{Simulation Setup} \label{sub:SS}

We consider the outdoor scenario `O1\textunderscore60' which is offered by the DeepMIMO dataset \cite{DeepMIMO}.
Two BSs are selected in the `O1\textunderscore60' scenario, which are BS 3 and 4 (acting as BS 1 and 2 in this paper\footnote{Without ambiguity, in this subsection, BS 1 and BS 3 are used interchangeably. Similarly, BS 2 and BS 4 are used interchangeably.}).
%
The parameters used for generating the channels are summarized in \tref{param}.
Based on these parameters, the pairwise BS to user channel vectors are generated, as well as the interference channel matrix between two BSs.
Besides, given that there are two BSs, each user in the selected user grid therefore has two channels that correspond to different BSs.
The user association (i.e., the user is served by which BS) is determined by comparing the channel strengths (i.e., the 2-norm of the channel vectors) between BS 1 and 2, and the user is communicating with the BS that has stronger channel.
For clarity, we denote the set of users (i.e., their channels) that are served by BS 1 as $\boldsymbol{\mathcal{H}}_1$, and the other users that are served by BS 2 as $\boldsymbol{\mathcal{H}}_2$.

\ifnum\x=0
{\clearpage}
\fi

\subsection{Evaluation Method} \label{sub:EVM}

During the evaluation, we assume that BS 1 has a codebook $\boldsymbol{\mathcal{W}}_1$ and BS 2 has a codebook $\boldsymbol{\mathcal{W}}_2$, regardless of whether they are learned codebooks or some pre-defined codebooks.
In the case when BS 1 is receiving and BS 2 is transmitting (the other case, i.e., BS 2 is receiving and BS 1 is transmitting, is identical), for any user of BS 1's, i.e., $\mathbf{h}\in\boldsymbol{\mathcal{H}}_1$, the beam used by BS 1 for serving this user, i.e., $\mathbf{w}$, is found through \eqref{beam-training-w}.
The average signal-to-interference ratio (SIR) performance achieved by this beam at this user, denoted as $\mathrm{SIR}_{\mathrm{avg}}(\mathbf{h},\mathbf{w})$, can be approximately calculated as below
\begin{equation}\label{i-level}
  \mathrm{SIR}_{\mathrm{avg}}(\mathbf{h}, \mathbf{w}) = \mathbb{E}_{\mathbf{f}\in\boldsymbol{\mathcal{W}}_2}\left[\frac{|\mathbf{w}^H\mathbf{h}|^2}{\left|\mathbf{w}^H\mathbf{H}\mathbf{f}\right|^2}
  \right] \approx \frac{1}{|\boldsymbol{\mathcal{W}}_2|}\sum_{\mathbf{f}\in\boldsymbol{\mathcal{W}}_2}\frac{|\mathbf{w}^H\mathbf{h}|^2}{\left|\mathbf{w}^H\mathbf{H}\mathbf{f}\right|^2},
\end{equation}
where we assume that the beams in BS 2's codebook have even chance to be selected.
It is worth pointing out that \eqref{i-level} is not the only metric for evaluating the interference suppression capability. For example, one can use the minimum SIR achieved by this beam at this user as its performance, i.e., $\mathrm{SIR}_{\mathrm{min}}(\mathbf{h}, \mathbf{w}) = \min_{\mathbf{f}\in\boldsymbol{\mathcal{W}}_2}\frac{|\mathbf{w}^H\mathbf{h}|^2}{\left|\mathbf{w}^H\mathbf{H}\mathbf{f}\right|^2}$.
This is because oftentimes, the strong interference can directly lead to the outage of the user.

\ifnum\x=0
{\clearpage}
\fi

\subsection{Numerical Results} \label{sub:NR-CB}
\highlight{
In this subsection, we present the numerical evaluation results of the proposed fully decentralized interference-aware beam codebook design approach.
}

\subsubsection{Hypothesis testing accuracy}

\begin{figure*}[t]
	\centering
	\subfigure[Random transmit beams]{ \includegraphics[width=.45\textwidth]{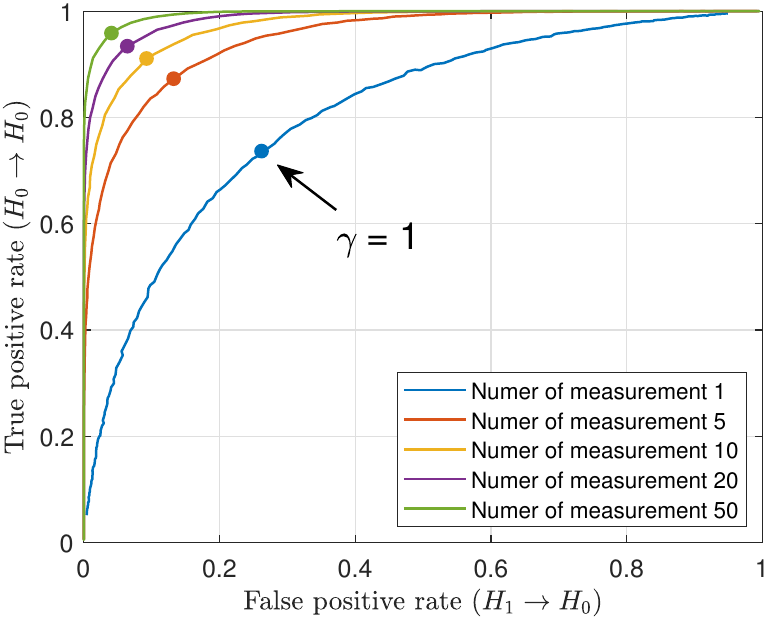} \label{roc:random} }
    \subfigure[DFT transmit beams]{ \includegraphics[width=.45\textwidth]{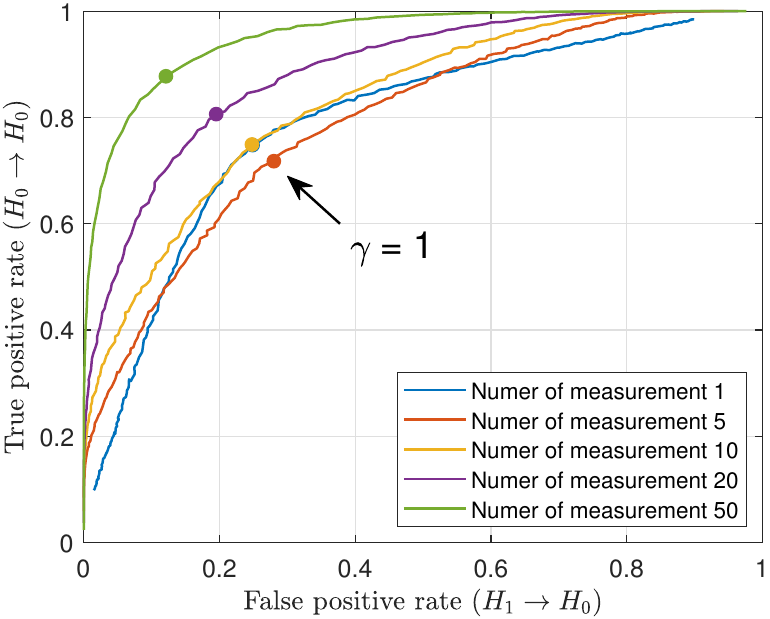} \label{roc:dft} }
    \caption{The ROC curves of the proposed decision rule with different number of measurements, where (a) shows the case when the other BS is using random transmit beams and (b) shows the case when the other BS is using a fixed DFT codebook.}
	\label{fig:roc}
\end{figure*}

We first evaluate the detection accuracy of the proposed decision rule, i.e., \eqref{dec-0} and \eqref{dec-1}, in determining the quality of two different beams in suppressing the interference signals.
Such capability is important as it provides the learning agent with robust information against the non-stationary interference.
To this end, we plot the receiver operating characteristic (ROC) performance curves of the proposed decision rule under different number of measurements.
To generate such curves, we re-write \eqref{dec-0} and \eqref{dec-1} in the following form
\begin{equation}\label{hypo-gamma}
  \frac{\mathbb{E}\left[\left|\mathbf{w}^H\mathbf{H}\mathbf{f}\right|^2\right]}{\mathbb{E}\left[\left|{\mathbf{w}^\prime}^H\mathbf{H}\mathbf{f}\right|^2\right]} \underset{H_1}{\overset{H_0}{\lessgtr}} \gamma,
\end{equation}
where $\gamma$ is the threshold of the decision rule (or detector).
From \eqref{dec-0} and \eqref{dec-1}, it implies that $\gamma = 1$.
\highlight{
Moreover, in the simulation, we define the true positive rate (TPR) as the probability of the proposed decision rule predicting $H_0$ when $H_0$ is indeed true, i.e., $\|\boldsymbol{\xi}\|_2^2 < \|\boldsymbol{\xi}^\prime\|_2^2$.
The false positive rate (FPR) is defined as the probability of the proposed decision rule predicting $H_0$ when $H_1$ is actually true, i.e., $\|\boldsymbol{\xi}\|_2^2 \ge \|\boldsymbol{\xi}^\prime\|_2^2$.
}
As can be seen, the larger the TPR and the smaller the FPR, i.e., more towards the upper left corner of the plot, the better the decision rule.

In \fref{roc:random}, we plot the performance of the proposed decision rule when BS 2 is transmitting signals with random beams.
For instance, the BS 2 could be also during a beam learning process and hence does not possess a fixed beam switch pattern.
As can be seen, with more measurements, i.e., larger $P$ in \eqref{int-seq}, the detection accuracy of the proposed solution is improved quite significantly.
This is as expected since the system can better combat the non-stationarity effect when acquiring more measurements and performing averaging.
Specifically, as can be observed in \fref{roc:random}, when the number of measurement is $10$, and a threshold of $\gamma=1$ is adopted, the system is able to achieve higher than $90\%$ TPR and lower than $10\%$ FPR.
Such accurate estimation of the quality of different beams can largely provide a smooth and consistent learning experience for the agent.

In \fref{roc:dft}, we plot the performance of the proposed decision rule when BS 2 is using a fixed beam codebook.
To be more specific, we assume that BS 2 adopts a 16-beam DFT codebook.
As can be seen, the proposed solution is still able to achieve satisfying performance, and with more measurements, the estimation accuracy is able to further improved.
However, it is not as good as random beam.
This empirically highlights that under the current estimation framework, random beams are more favorable.
However, it does not imply that the DFT beams are more difficult.
For instance, in case when BS 1 knows that BS 2 will use a fixed beam codebook with finite number of beams, there might exist better estimation methods.

\ifnum\x=0
{\clearpage}
\fi

\subsubsection{SIR map}

\begin{figure*}[t]
	\centering
    \subfigure[Outdoor scenario]{ \includegraphics[width=0.36\textwidth]{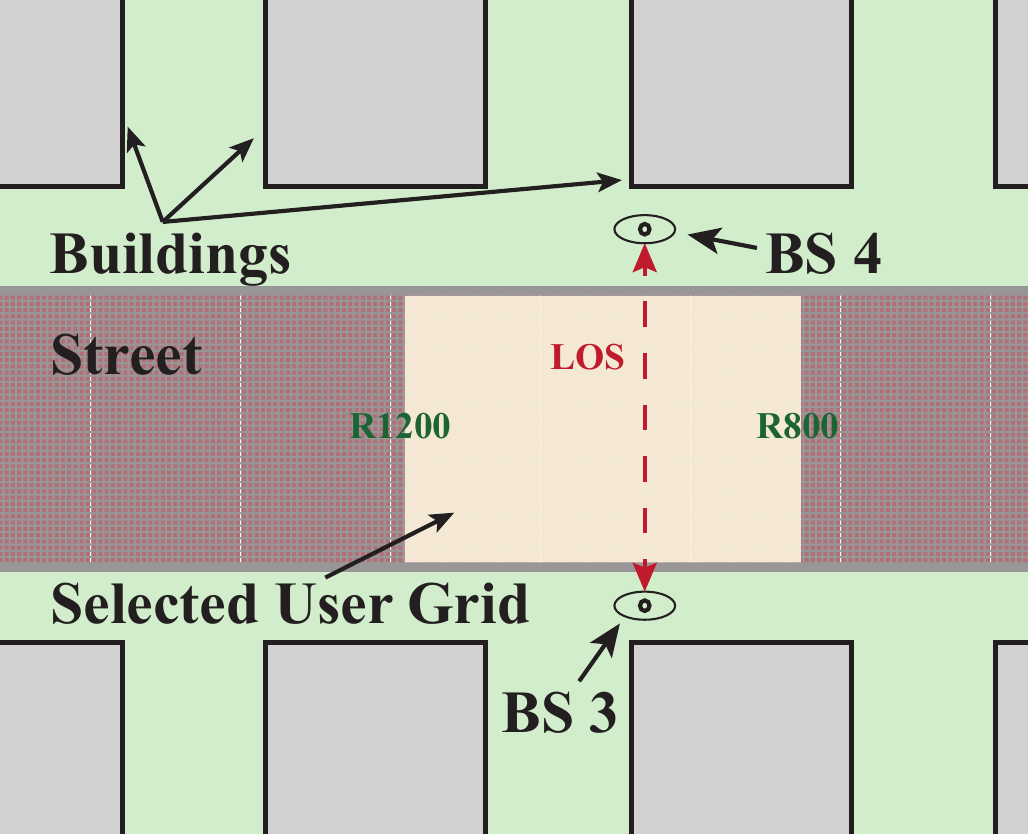} \label{sir-map:0} }
    \subfigure[Beamsteering codebooks]{ \includegraphics[width=0.273\textwidth]{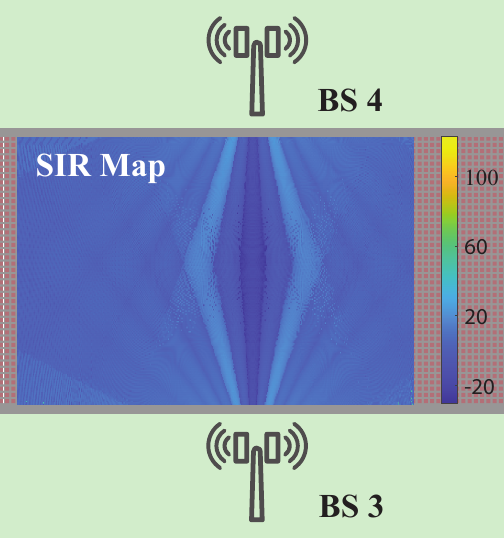} \label{sir-map:a} }
	\subfigure[Learned codebooks]{ \includegraphics[width=0.273\textwidth]{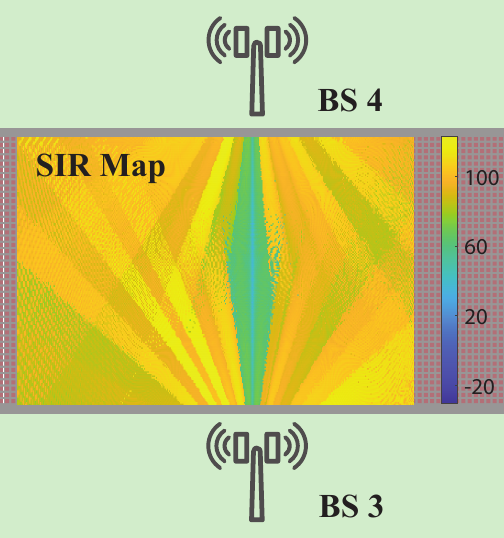} \label{sir-map:b} }
	\caption{The simulation results of the proposed beam codebook learning solution, where (a) depicts the considered outdoor communication scenario, (b) shows the achieved SIR map (in decibel) when both BSs using the beamsteering codebooks, and (c) shows the achieved SIR map (in decibel) when both of them using the learned codebooks.}
	\label{sir-map}
\end{figure*}

In \fref{sir-map:0}, we depict the adopted outdoor communication scenario, where there are two BSs and a grid of users in between of them. As illustrated in the figure, there exists a direct LOS link between these two BSs, which incurs strong interference.
We first evaluate the SIR performance when both BSs employ a 16-beam \highlight{beamsteering codebook} (which is common in the current 5G NR systems), and the achieved SIR at each user is calculated based on \eqref{i-level}.
As can be observed from \fref{sir-map:a}, \textbf{the majority of the users are experiencing around $0$ dB SIR,} which cannot meet the stringent requirements in most of the real applications.
%
Despite the narrow beam patterns of the beamsteering beams, the adoption of such codebooks still produces quite unsatisfactory performance.
%
This is because given that the interference channel has much higher power then the users' channels, even side-lobes of a beam, either transmit beam of the interferer or the combining beam of the receiver, can cause rather high interference level.
By contrast, as shown in \fref{sir-map:b}, by deploying the learned codebooks, the SIR performance of the considered user grid is significantly improved, \textbf{yielding over $80$ dB SIR on most of the users.}
It is worth noting that only the users that are exactly at the line segment connecting these two BSs have relatively worse SIR performance.
This is because their channel vectors have almost the same spatial directions as the interfering direction, leading to unavoidable interference, especially when the interfering BS also serves its own users along these directions.
One way to address such problem, therefore, is to somehow form a consensus between these BSs such that they can have better user scheduling.

\ifnum\x=0
{\clearpage}
\fi

\subsubsection{Codebook patterns}

\begin{figure*}[t]
	\centering
	\subfigure[]{ \includegraphics[width=.31\textwidth]{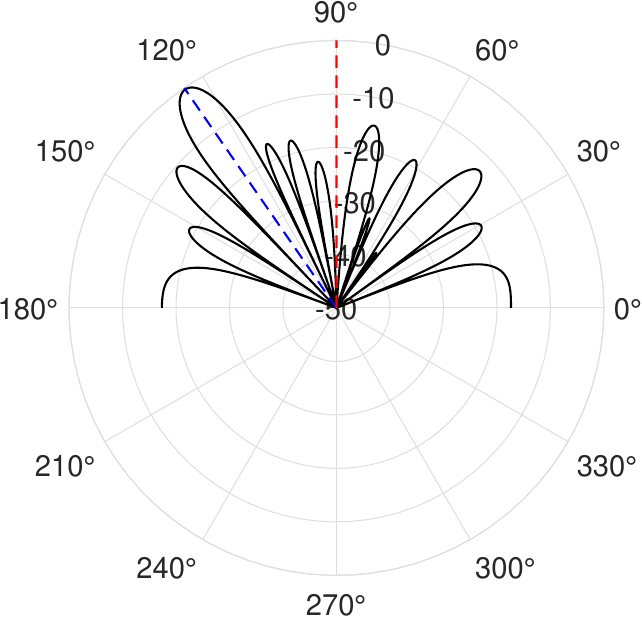} \label{bs3:a} }
    \subfigure[]{ \includegraphics[width=.31\textwidth]{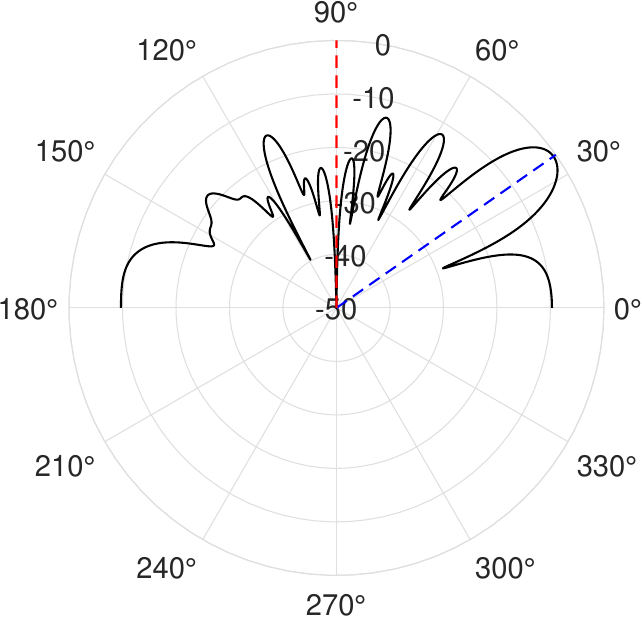} \label{bs3:b} }
    \subfigure[]{ \includegraphics[width=.31\textwidth]{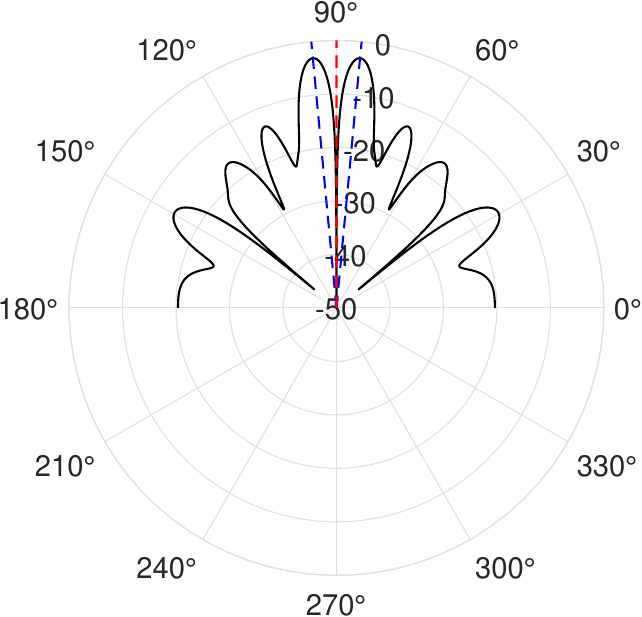} \label{bs3:d} }
    \caption{The selected beam patterns from the learned 16-beam codebook of BS 3, where the dashed blue line indicating the direction of the main lobe of the beam and the dashed red line indicating the direction of the interference.}
	\label{fig:bs3-pat}
\end{figure*}

\begin{figure*}[t]
	\centering
	\subfigure[]{ \includegraphics[width=.31\textwidth]{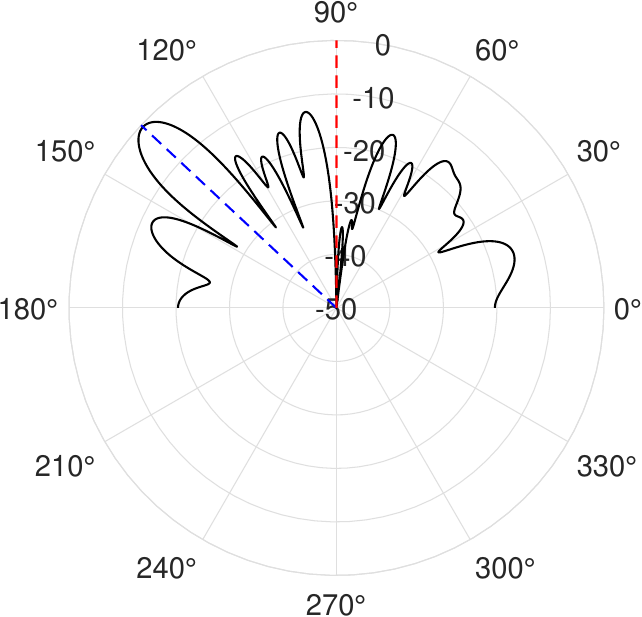} \label{bs4:a} }
    \subfigure[]{ \includegraphics[width=.31\textwidth]{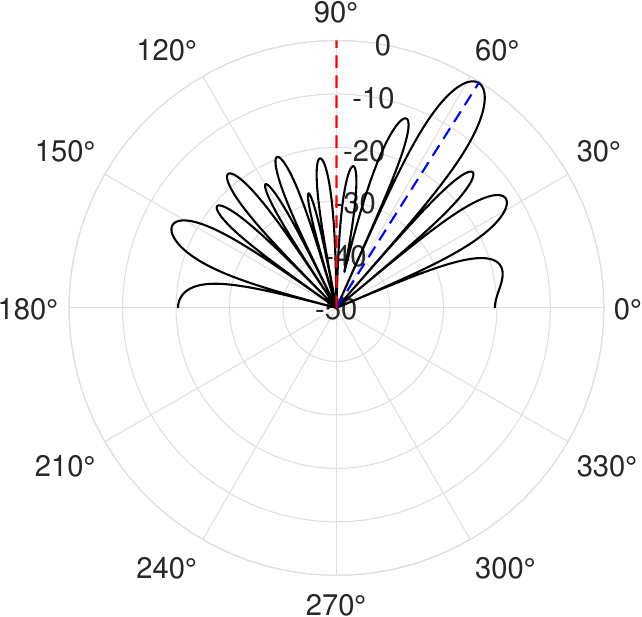} \label{bs4:b} }
    \subfigure[]{ \includegraphics[width=.31\textwidth]{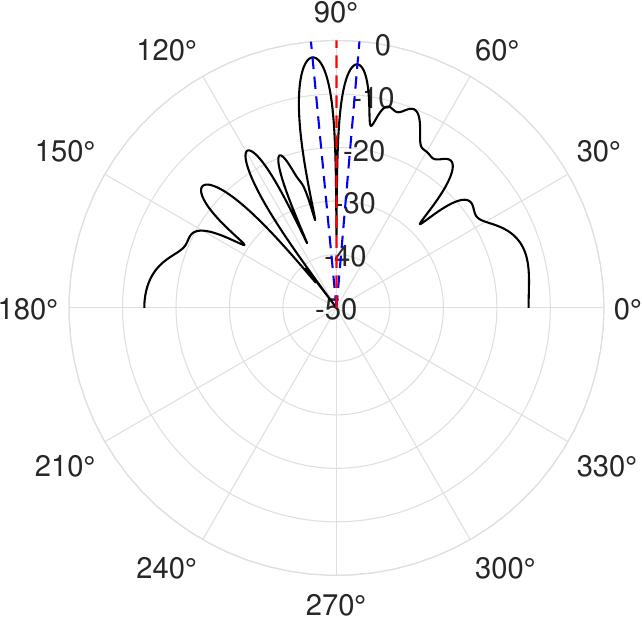} \label{bs4:d} }
    \caption{The selected beam patterns from the learned 16-beam codebook of BS 4, where the dashed blue line indicating the direction of the main lobe of the beam and the dashed red line indicating the direction of the interference.}
	\label{fig:bs4-pat}
\end{figure*}

\highlight{
To gain some insights of how the learned codebooks are able to outperform the beamsteering codebooks to such a great extent, we also plot the learned beam patterns in these codebooks.
}
As can be seen in \fref{fig:bs3-pat} and \fref{fig:bs4-pat}, where we present few selected beam patterns from each codebook, all the beams have very deep nulls towards the direction of interference, which almost eliminates the undesired signal from that direction.
Besides, it is interesting to observe that the beams for serving the users at the vicinity of the middle line segment that connects the two BSs (\fref{bs3:d} and \fref{bs4:d}) break into two parts.
This somehow explains the SIR map obtained in \fref{sir-map:b}.
Specifically, the design of such beam needs to strike a delicate balance between better receiving the desired signals from the users and suppressing the undesired interference signals.
This is unlike the other beams where such trade-off is not that conflictive.

Furthermore, it is worth emphasizing that such interference-aware beam codebooks are designed under the following critical conditions:
(i) The BSs do not know any explicit channel knowledge of both their served users and the interference channel between each other;
(ii) The BSs are operating in an non-cooperative and fully decentralized way with no information sharing between them.
\textbf{This makes the solution implementation friendly and can be used by asynchronous wireless networks that are possibly controlled by different operators.}
\highlight{
Finally, as the learned beam codebook for each BS will be used for both transmitting and receiving signals, the algorithm essentially encourages all the BSs to avoid receiving, and hence transmitting, from the interfering directions, which is helpful for both individual as well as for the other BSs.
}

\ifnum\x=0
{\clearpage}
\fi

\section{Conclusions and Discussions} \label{sec:Con}

In this paper, we generalize the interference-aware codebook learning with multiple base stations, and consider the challenging scenario when there is no information exchange allowed between the different base stations.
To tackle the non-stationarity caused by the operations of the adjacent base stations, the power measurement average is employed to estimate the interference nulling performance of a particular beam configuration, based on which a decision rule is provided. Furthermore, we theoretically justify the adoption of such estimator, and prove that the provided decision rule is sufficient in determining the quality of a beam in large antenna array system with limited multi-paths.
This lays the foundation for the proposed reward function in the multi-agent reinforcement learning, which is essential in fully decoupling the learning of the different agents running at each base station.
Simulation results show that the developed solution is capable of learning well-shaped codebook patterns for different base stations that significantly suppress the interference without information exchange and relying only on power measurements. For future work, it is interesting to investigate how digital twins \cite{digitaltwin_Ahmed, digitalTwin_Jiang} could be leveraged to further reduce the data collection/model training overhead. It will also be interesting to explore possible enhancements when other sensing modalities are considered.

\appendix
\section{}
\subsection{Proof of Proposition 1} \label{append-a}

We first examine the structure of $\boldsymbol{\Sigma}$, where by definition
\begin{equation}\label{cov-matrix}
  \boldsymbol{\Sigma} = \mathbb{E}[\mathbf{z}\mathbf{z}^H] = \mathbb{E}\left[
                                                               \begin{array}{cccc}
                                                                 \mathbf{a}_{\mathrm{t},1}^H \mathbf{f}\hspace{0.5pt}\mathbf{f}^H \mathbf{a}_{\mathrm{t},1} & \mathbf{a}_{\mathrm{t},1}^H \mathbf{f}\hspace{0.5pt}\mathbf{f}^H \mathbf{a}_{\mathrm{t},2} & \cdots & \mathbf{a}_{\mathrm{t},1}^H \mathbf{f}\hspace{0.5pt}\mathbf{f}^H \mathbf{a}_{\mathrm{t},L} \\
                                                                 \mathbf{a}_{\mathrm{t},2}^H \mathbf{f}\hspace{0.5pt}\mathbf{f}^H \mathbf{a}_{\mathrm{t},1} & \mathbf{a}_{\mathrm{t},2}^H \mathbf{f}\hspace{0.5pt}\mathbf{f}^H \mathbf{a}_{\mathrm{t},2} & \cdots & \mathbf{a}_{\mathrm{t},2}^H \mathbf{f}\hspace{0.5pt}\mathbf{f}^H \mathbf{a}_{\mathrm{t},L} \\
                                                                 \vdots & \vdots & \ddots & \vdots \\
                                                                 \mathbf{a}_{\mathrm{t},L}^H \mathbf{f}\hspace{0.5pt}\mathbf{f}^H \mathbf{a}_{\mathrm{t},1} & \mathbf{a}_{\mathrm{t},L}^H \mathbf{f}\hspace{0.5pt}\mathbf{f}^H \mathbf{a}_{\mathrm{t},2} & \cdots & \mathbf{a}_{\mathrm{t},L}^H \mathbf{f}\hspace{0.5pt}\mathbf{f}^H \mathbf{a}_{\mathrm{t},L} \\
                                                               \end{array}
                                                             \right],
\end{equation}
in which we use $\mathbf{a}_{\mathrm{t},\ell}$ to denote $\mathbf{a}_\mathrm{t}(\phi_{\mathrm{t}, \ell}, \vartheta_{\mathrm{t}, \ell})$ to simplify the expression.
As each $\mathbf{a}_{\mathrm{t},\ell}, \forall \ell$ is essentially an unknown but deterministic quantity, and we notice that
\begin{equation}\label{f-expect}
  \mathbb{E}[\mathbf{f}\hspace{0.5pt}\mathbf{f}^H] = \left[
                                                      \begin{array}{cccc}
                                                        \frac{1}{M}\mathbb{E}[e^{j\theta_1}e^{-j\theta_1}] & \frac{1}{M}\mathbb{E}[e^{j\theta_1}e^{-j\theta_2}] & \cdots & \frac{1}{M}\mathbb{E}[e^{j\theta_1}e^{-j\theta_M}] \\
                                                        \frac{1}{M}\mathbb{E}[e^{j\theta_2}e^{-j\theta_1}] & \frac{1}{M}\mathbb{E}[e^{j\theta_2}e^{-j\theta_2}] & \cdots & \frac{1}{M}\mathbb{E}[e^{j\theta_2}e^{-j\theta_M}] \\
                                                        \vdots & \vdots & \ddots & \vdots \\
                                                        \frac{1}{M}\mathbb{E}[e^{j\theta_M}e^{-j\theta_1}] & \frac{1}{M}\mathbb{E}[e^{j\theta_M}e^{-j\theta_2}] & \cdots & \frac{1}{M}\mathbb{E}[e^{j\theta_M}e^{-j\theta_M}] \\
                                                      \end{array}
                                                    \right] = \frac{1}{M}\mathbf{I}_{M\times M},
\end{equation}
we have \eqref{cov-matrix} simplified to
\begin{align}\label{cov-matrix-simp}
  \boldsymbol{\Sigma} = & \frac{1}{M}\left[
                                     \begin{array}{cccc}
                                       1 & \mathbf{a}_{\mathrm{t},1}^H\mathbf{a}_{\mathrm{t},2} & \cdots & \mathbf{a}_{\mathrm{t},1}^H\mathbf{a}_{\mathrm{t},L} \\
                                       \mathbf{a}_{\mathrm{t},2}^H\mathbf{a}_{\mathrm{t},1} & 1 & \cdots & \mathbf{a}_{\mathrm{t},2}^H\mathbf{a}_{\mathrm{t},L} \\
                                       \vdots & \vdots & \ddots & \vdots \\
                                       \mathbf{a}_{\mathrm{t},L}^H\mathbf{a}_{\mathrm{t},1} & \mathbf{a}_{\mathrm{t},L}^H\mathbf{a}_{\mathrm{t},2} & \cdots & 1 \\
                                     \end{array}
                                   \right], \\
  = & \frac{1}{M} \left(\mathbf{I}_{L\times L} + \left[
                                     \begin{array}{cccc}
                                       0 & \mathbf{a}_{\mathrm{t},1}^H\mathbf{a}_{\mathrm{t},2} & \cdots & \mathbf{a}_{\mathrm{t},1}^H\mathbf{a}_{\mathrm{t},L} \\
                                       \mathbf{a}_{\mathrm{t},2}^H\mathbf{a}_{\mathrm{t},1} & 0 & \cdots & \mathbf{a}_{\mathrm{t},2}^H\mathbf{a}_{\mathrm{t},L} \\
                                       \vdots & \vdots & \ddots & \vdots \\
                                       \mathbf{a}_{\mathrm{t},L}^H\mathbf{a}_{\mathrm{t},1} & \mathbf{a}_{\mathrm{t},L}^H\mathbf{a}_{\mathrm{t},2} & \cdots & 0 \\
                                     \end{array}
                                   \right] \right), \label{cov-2} \\
  = & \frac{1}{M} \left(\mathbf{I}_{L\times L} + \boldsymbol{\Pi}\right). \label{sigma-decomp}
\end{align}
%
Notice that $\boldsymbol{\Sigma}$ can be viewed as the identity matrix perturbed by $\boldsymbol{\Pi}$. Therefore, by leveraging the Weyl's inequality
\begin{equation}\label{weyl}
  \max_k \left|\lambda_k(\mathbf{I} + \boldsymbol{\Pi}) - \lambda_k(\mathbf{I})\right| \le \left\|\boldsymbol{\Pi}\right\|_2,
\end{equation}
where obviously we have $\lambda_k(\mathbf{I})=1, \forall k$, and by noticing that $\lambda_k(\mathbf{I} + \boldsymbol{\Pi})>0, \forall k$ (as $\boldsymbol{\Sigma}\succ 0$),
%
%
we have
\begin{equation}\label{eigenvalues}
  0 < \lambda_L(\mathbf{I} + \boldsymbol{\Pi}) \le \lambda_k(\mathbf{I} + \boldsymbol{\Pi}) \le 1 + \left\|\boldsymbol{\Pi}\right\|_2, ~ \forall k\in\{1,\dots,L\}.
\end{equation}
Therefore, we express $\boldsymbol{\Sigma}$ by using its spectral decomposition, i.e., $\boldsymbol{\Sigma} = \mathbf{U}\boldsymbol{\Lambda}\mathbf{U}^H$, where $\mathbf{U}$ is an unitary matrix and $\boldsymbol{\Lambda}$ contains the eigenvalues of $\boldsymbol{\Sigma}$, in a decreasing order, i.e., $\boldsymbol{\Lambda} = \mathrm{diag}\left(\lambda_1(\boldsymbol{\Sigma}), \lambda_2(\boldsymbol{\Sigma}), \dots, \lambda_L(\boldsymbol{\Sigma})\right)$.

Now, we examine the quadratic term in \eqref{varterm}, by using the Rayleigh quotient, that is
\begin{equation}\label{rayleigh}
  0 < \frac{1}{M}\lambda_L(\mathbf{I} + \boldsymbol{\Pi}) = \lambda_L(\boldsymbol{\Sigma}) \le \frac{\tilde{\boldsymbol{\xi}}^H\boldsymbol{\Sigma}\tilde{\boldsymbol{\xi}}}{\tilde{\boldsymbol{\xi}}^H\tilde{\boldsymbol{\xi}}} \le \lambda_1(\boldsymbol{\Sigma}) \le \frac{1}{M}\left(1 + \left\|\boldsymbol{\Pi}\right\|_2\right),
\end{equation}
which implies that
\begin{equation}\label{rayleigh-2}
  \frac{1}{M}\lambda_L(\mathbf{I} + \boldsymbol{\Pi})\|\tilde{\boldsymbol{\xi}}\|_2^2 \le \tilde{\boldsymbol{\xi}}^H\boldsymbol{\Sigma}\tilde{\boldsymbol{\xi}} \le \frac{1}{M}\left(1 + \left\|\boldsymbol{\Pi}\right\|_2\right)\|\tilde{\boldsymbol{\xi}}\|_2^2.
\end{equation}
This implies that, if there exists another $\tilde{\boldsymbol{\xi}^\prime}$, such that \eqref{condition} is satisfied,
then, it is guaranteed that $\mathbb{E}\left[\mathbf{z}^H\mathbf{A}\mathbf{z}\right]<\mathbb{E}\left[\mathbf{z}^H\mathbf{A}^\prime\mathbf{z}\right]$.

\subsection{Proof of Proposition 2} \label{append-b}

As $\left\|\boldsymbol{\Pi}\right\|_2\ge 0$, we have
\begin{align}
  \mathbb{P}\left[\left|1-\frac{1-\left\|\boldsymbol{\Pi}\right\|_2}{1+\left\|\boldsymbol{\Pi}\right\|_2}\right| > \epsilon\right] = &~ \mathbb{P}\left[\frac{2\left\|\boldsymbol{\Pi}\right\|_2}{1+\left\|\boldsymbol{\Pi}\right\|_2} > \epsilon\right], \\
  = &~ \mathbb{P}\left[\left\|\boldsymbol{\Pi}\right\|_2 > \frac{\epsilon}{2-\epsilon}\right], \\
  < &~ \mathbb{P}\left[\left\|\boldsymbol{\Pi}\right\|_2 > \frac{\epsilon}{2}\right], \\
  \overset{(a)}{\le} &~ \mathbb{P}\left[\left\|\boldsymbol{\Pi}\right\|_1 > \frac{\epsilon}{2}\right],
\end{align}
where (a) is due to $\left\|\boldsymbol{\Pi}\right\|_2 \le \left\|\boldsymbol{\Pi}\right\|_1$.
Moreover, we have
\begin{align}
   & \lim_{\frac{M}{L}\rightarrow +\infty}\left\|\boldsymbol{\Pi}\right\|_1 \\
  = & \lim_{\frac{M}{L}\rightarrow +\infty}\max_{\ell}\left(\sum_{\ell^\prime\ne\ell}\left|\mathbf{a}_\mathrm{t}(\phi_{\mathrm{t}, \ell^\prime}, \vartheta_{\mathrm{t}, \ell^\prime})^H\mathbf{a}_\mathrm{t}(\phi_{\mathrm{t}, \ell}, \vartheta_{\mathrm{t}, \ell})\right|^2\right), \\
  = &~ \max_{\ell}\sum_{\ell^\prime\ne\ell}\lim_{\frac{M}{L}\rightarrow +\infty}\left|\mathbf{a}_\mathrm{t}(\phi_{\mathrm{t}, \ell^\prime}, \vartheta_{\mathrm{t}, \ell^\prime})^H\mathbf{a}_\mathrm{t}(\phi_{\mathrm{t}, \ell}, \vartheta_{\mathrm{t}, \ell})\right|^2, \\
  \overset{(a)}{=} &~ 0, \label{ortho-norm}
\end{align}
where the proof of (a) can be found in the proof of Corollary 2 in \cite{Ayach2012capacity}.
Therefore, we have
\begin{equation}
  \lim_{\frac{M}{L}\rightarrow +\infty}\mathbb{P}\left[\left\|\boldsymbol{\Pi}\right\|_1 > \frac{\epsilon}{2}\right] = 0,
\end{equation}
and hence
\begin{equation}
  \lim_{\frac{M}{L}\rightarrow +\infty}\mathbb{P}\left[\left|1-\frac{1-\left\|\boldsymbol{\Pi}\right\|_2}{1+\left\|\boldsymbol{\Pi}\right\|_2}\right| > \epsilon\right] = 0.
\end{equation}

\linespread{1.3}
\bibliographystyle{IEEEtran}

\end{document}

%% file: input.tex
\usepackage{amsfonts}
\usepackage{times}
\usepackage{graphicx}
\usepackage{latexsym}
\usepackage{dsfont}
\usepackage{amssymb}
\usepackage{amsmath}
\usepackage{cite}
\usepackage{verbatim}
\usepackage{subfigure}
\newtheorem{theorem}{Theorem}

\newtheorem{corollary}[theorem]{Corollary}

\def\bb0{{\mathbb{0}}}

\def\bb{{\mathbf{b}}}

\def\b0{{\mathbf{0}}}

\def\sf0{{\mathsf{0}}}

